\documentclass[authorversion,sigconf]{acmart}
\AtBeginDocument{%
  \providecommand\BibTeX{{%
    \normalfont B\kern-0.5em{\scshape i\kern-0.25em b}\kern-0.8em\TeX}}}

\setcopyright{acmcopyright}

\copyrightyear{2024}
\acmYear{2024}
\setcopyright{acmlicensed}\acmConference[CHI '24]{Proceedings of the CHI
Conference on Human Factors in Computing Systems}{May 11--16,
2024}{Honolulu, HI, USA}
\acmBooktitle{Proceedings of the CHI Conference on Human Factors in
Computing Systems (CHI '24), May 11--16, 2024, Honolulu, HI, USA}
\acmDOI{10.1145/3613904.3642711}
\acmISBN{979-8-4007-0330-0/24/05}

\usepackage{xspace}
\newcommand{\ie}{i.e.{\xspace}}
\newcommand{\eg}{e.g.{\xspace}}

\newcommand{\rev}[1]{#1}
\newcommand{\qt}[1]{\textit{``#1''}}
\newcommand{\pqt}[2]{\textit{``#1''}{\,}{\small-#2}}

\newcommand{\etal}{{\xspace}et~al.{\xspace}}

\usepackage{amsmath}
\usepackage{graphicx}
\usepackage{subcaption}

\newcommand{\sysname}{Piet}
\begin{document}

%%
%% The "title" command has an optional parameter,
%% allowing the author to define a "short title" to be used in page headers.
\title{Piet: Facilitating Color Authoring for Motion Graphics Video}

%%
%% The "author" command and its associated commands are used to define
%% the authors and their affiliations.
%% Of note is the shared affiliation of the first two authors, and the
%% "authornote" and "authornotemark" commands
%% used to denote shared contribution to the research.
% \author{Ben Trovato}
% \authornote{Both authors contributed equally to this research.}
% \email{trovato@corporation.com}
% \orcid{1234-5678-9012}
% \author{G.K.M. Tobin}
% \authornotemark[1]
% \email{webmaster@marysville-ohio.com}
% \affiliation{%
%   \institution{Institute for Clarity in Documentation}
%   \streetaddress{P.O. Box 1212}
%   \city{Dublin}
%   \state{Ohio}
%   \country{USA}
%   \postcode{43017-6221}
% }

\author{Xinyu Shi}
\affiliation{%
  \institution{University of Waterloo}
  \city{Waterloo}
  \state{ON}
  \country{Canada}
}
\email{xinyu.shi@uwaterloo.ca}
% \authornote{Work done during an internship at Microsoft Research Asia.}

\author{Yinghou Wang}
\affiliation{%
  \institution{Harvard University}
  \city{Cambridge}
  \state{MA}
  \country{United States}
}
\email{yinghouwang@gsd.harvard.edu}

\author{Yun Wang}
\affiliation{%
  \institution{Microsoft Research Asia}
  \city{Beijing}
  % \state{ON}
  \country{China}
}
\email{wangyun@microsoft.com}

\author{Jian Zhao}
\affiliation{%
  \institution{University of Waterloo}
  \city{Waterloo}
  \state{ON}
  \country{Canada}
}
\email{jianzhao@uwaterloo.ca}

%%
%% By default, the full list of authors will be used in the page
%% headers. Often, this list is too long, and will overlap
%% other information printed in the page headers. This command allows
%% the author to define a more concise list
%% of authors' names for this purpose.
\renewcommand{\shortauthors}{Xinyu Shi, et al.}

%%
%% The abstract is a short summary of the work to be presented in the
%% article.
\begin{abstract}
Motion graphic (MG) videos are effective and compelling for presenting complex concepts through animated visuals; and colors are important to convey desired emotions, maintain visual continuity, and signal narrative transitions.
However, current video color authoring workflows are fragmented, lacking contextual previews, hindering rapid theme adjustments, and not aligning with designers' progressive authoring flows.
To bridge this gap, we introduce \sysname{}, the first tool tailored for MG video color authoring. 
\sysname{} features an interactive palette to visually represent color distributions, support controllable focus levels, and enable quick theme probing via grouped color shifts.
We interviewed 6 domain experts to identify the frustrations in current tools and inform the design of \sysname{}.
An in-lab user study with 13 expert designers showed that \sysname{} effectively simplified the MG video color authoring and reduced the friction in creative color theme exploration.
\end{abstract}

\begin{teaserfigure}
    \centering
    \includegraphics[width=1.\linewidth]{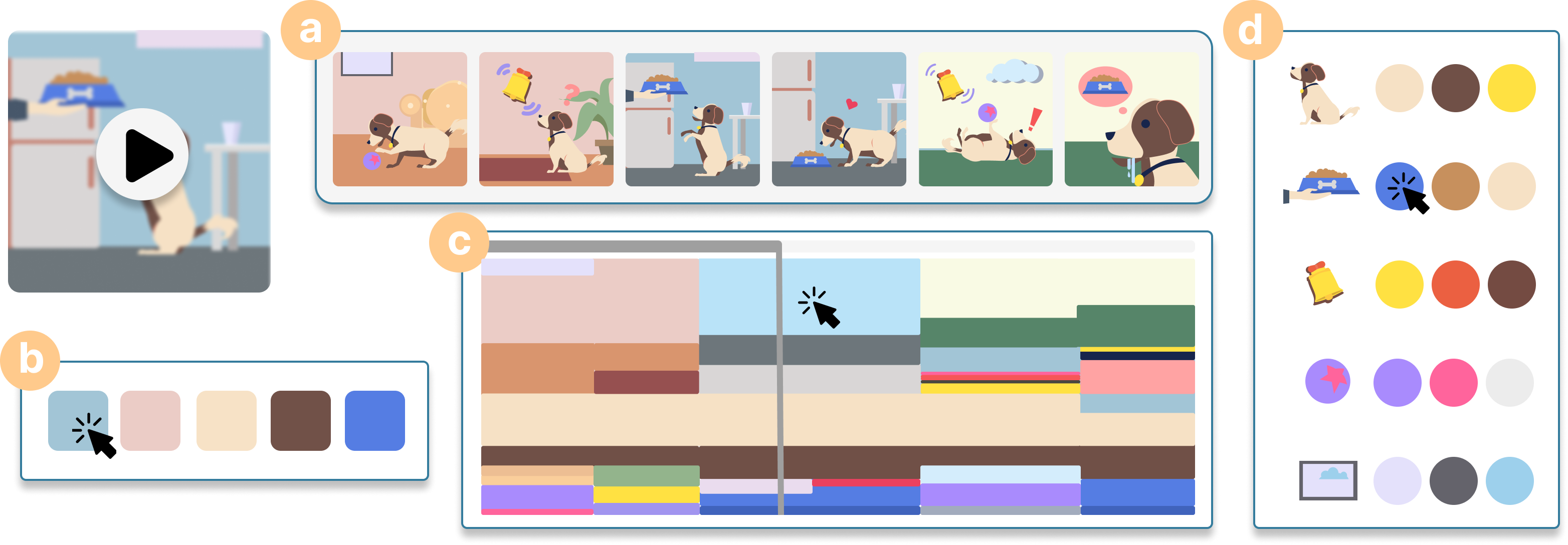}
    \vspace{-6mm}
    \caption{\sysname{} is a color authoring tool tailored to motion graphics videos such as the animated visuals shown in (a). \sysname{} provides different interactive color representation views: (b) \textit{Video Theme} summarizes the dominant colors; (c) \textit{Scene Palette} describes the detailed color distribution throughout the video timeline; and (d) \textit{Element List} shows individual colors for each element.
    }
    \label{fig:teaser}
    \Description{The teaser figure provides a detailed overview of \sysname{}, a tailored tool designed for color management in motion graphics videos. This figure is divided into four parts: part (a) showcases an example of motion graphics videos explaining the Pavlov's dog experiment; part (b) shows five dominant colors in the example video; part (c) illustrates the detailed color distribution across the video's timeline, providing a comprehensive view of how colors vary and transition throughout the video; part (d) lists individual colors associated with each element in the video.}
\end{teaserfigure}

%%
%% The code below is generated by the tool at http://dl.acm.org/ccs.cfm.
%% Please copy and paste the code instead of the example below.
%%



%%
%% Keywords. The author(s) should pick words that accurately describe
%% the work being presented. Separate the keywords with commas.

%%
%% This command processes the author and affiliation and title
%% information and builds the first part of the formatted document.
\maketitle
\section{Introduction}

Motion graphic (MG) videos are short animated videos crafted to engage audiences in a logically connected narrative and facilitate content comprehension \cite{amini2015understanding, nordmark2012mobile}.
By delivering the visual, animated, and auditory elements in an organized and appealing manner, these videos make it effective and memorable to simplify complex concepts or narrate a brand story  \cite{jahanlou2022katika, jahanlou2020challenges, cheng2022investigating}.
The rise of online video platforms motivates both institutions and individual creators to leverage such videos for enhanced outreach. 
As a result, these videos are widely popular in various fields such as education and marketing \cite{sarinastiti2016skill, ccifcci2016effects, pate2020creating}.
For instance, Figure~\ref{fig:teaser}(a) shows frames from an explainer MG video introducing Pavlov's dog experiment. 

% In contrast to entertainment-focused MG videos, which emphasize viewer engagement through imaginative and fantastic visuals that may not always tie directly to the narration, explainer MG videos tightly bind the visuals to the narration, aiding easy and straightforward comprehension of content.

Color is a powerful visual language to shape the effectiveness of MG video.
As evidenced in static graphic designs, color evokes emotions, improves visual harmony, and directs viewers' attention \cite{eiseman2000pantone, barnard2013graphic, frascara2004communication}. 
Particular to MG videos, color is also essential to maintain \emph{visual continuity}. 
Across a series, consistent theme colors establish the brand identity, while within individual videos, adherence to a color theme ensures both captivating visuals and lasting impressions. 
Furthermore, colors can signal \emph{narrative transitions}: a sudden change signifies a plot shift, while a subtle gradient indicates seamless scene progression. 
Such complex roles of color demand careful design to achieve desired visual effects,
often causing designers to invest considerable time in iterating and repeatedly modifying colors.

Despite the extensive studies in HCI and media have investigated approaches to animation authoring \cite{jahanlou2022katika, kazi2014kitty}, exploration of visual storytelling \cite{lee2015more, porteous2010interactive}, and the broader difficulties faced in motion graphics creation \cite{jahanlou2020challenges}, the specific process and challenges of color design in MG videos remain largely unexplored. 
Despite the wealth of research addressing the challenges and opportunities in static graphics colorization \cite{jalal2015color, shugrina2020nonlinear, shugrina2017playful, chevalier2012histomages}, the knowledge gap in MG video color design and authoring is evident. 
Currently, there is a limited understanding of %the causes behind the time-consuming nature of 
the color authoring process in MG videos and the specific challenges designers face during this process. 
Aiming to bridge this gap, we conducted a formative study with six domain experts.
Our objective is to systematically explore how they perform color design in MG video creation, pinpoint what challenges they encounter, and identify design opportunities to streamline the color authoring experience in this field.

% \jian{IMPORTANT: I found we talked too much about the results of the formative study but not the motivation of doing the study. Here you should describe some literature that did not fully investigate this problem. You should also significantly shorten the three challenge paragraphs and/or move some content to the results of the formative study (Sec 3.3). }
% \fixme{The findings reveal the following three main challenges, primarily stemming from the current fragmented workflows and the deficiency of current tools that are not adequately adapted to video color authoring.} 

Our formative study identified three primary challenges in MG video color authoring. 
First, the current workflow separates the element creation and coloring from animation, depriving designers of immediate contextual previews to gauge the impact of color alterations on the overall video. 
Typically, designers use Adobe Illustrator\footnote{\url{https://www.adobe.com/products/illustrator.html} \label{AI}} for coloring and then switch to After Effects\footnote{\url{https://www.adobe.com/products/aftereffects.html} \label{AE}} for animation. 
Making color modifications often means frequently switching between these tools to ensure the edits do not disrupt the visual continuity and narrative flow, thus leading to a disruptive and time-consuming experience.
Second, current tools lack the support for joint adjustments for multiple elements in different frames, \eg, linearly shifting the saturation of a group of colors to be more vibrant while preserving their distinct hues.
This limitation prevents designers from creatively exploring various color theme alternations and makes post-animation color theme changes difficult. 
Lastly, existing tools lean more towards an object-driven methodology, forcing designers to always start from the most details for each object, contrasting to the fact that many designers prefer a top-down, progressive approach to color editing. 
The top-down strategy fosters visual continuity by prioritizing the dominant color adjustments and holding off on minor refinements, and thus designers can maintain a big picture, instead of being distracted from isolated object hues.
%By holding off on minor detail color refinements, designers can maintain a broader perspective on color distribution instead of being distracted from isolated object hues. 
%Despite its clear benefits, this approach finds little support in current tools, contrasting to the philosophy that many designers aspire to follow in their creative processes.

The challenges outlined above make MG colorization a time-consuming and complex task. 
Even though there are several tools available, including mainstream ones like Adobe Illustrator\textsuperscript{\ref{AI}} and After Effects\textsuperscript{\ref{AE}}, and lighter alternatives such as the Lottie editor\footnote{\url{https://lottiefiles.com/lottie-editor} \label{lottie}} and Figma\footnote{\url{https://www.figma.com} \label{Figma}}, none has fully addressed pressing difficulties of color authoring. 
For example, while the Lottie editor offers a feature for quick color palette selection in MG projects, it restricts users to a limited, predefined five-color palette, offering no room for detailed adjustments. 
Platforms like Figma\textsuperscript{\ref{Figma}}, Animaker\textsuperscript{\ref{Animaker}}, and Vimeo\textsuperscript{\ref{Vimeo}}, while easing the animation creation process to some degree, do not have features that specifically tailored to designers' needs for the color authoring. %, leaving a gap in intuitive and efficient color authoring experience.

% \jian{IMPORTANT: after you shorten the challenge paragraphs, here you should review the key systems developed in the past that do not fully address the challenges. This will motivate the development of the system.}

% To address these, we first conducted a formative study with six expert motion graphics designers to understand their existing workflows and the challenges they encounter in color authoring. 
To address these challenges, we introduce \sysname{}, an interactive tool for MG video color authoring with a more integrated workflow, multi-level color adjustment, and progressive editing (\autoref{fig:teaser}). 
% \jian{see if this statement is good}
\sysname{} combines color authoring and video playback, and proposes an interactive video palette that offers a visual abstraction for color distributions across the video. 
% \jian{maybe add a sentence to comment on the visual design of the color palette, e.g., Mosaic}
As shown in Figure~\ref{fig:teaser}(c), the video palette exhibits a visual style of dynamic Mosaics, illustrating the dynamic fluctuations in colors and their proportions throughout the video timeline.
Designers can directly manipulate this palette for multi-faceted color adjustments, with the flexibility to zoom in and out to control their desired focus levels.
Specifically, \sysname{} supports context-aware, group-wise, and multi-level color modifications spanning video, scene, and object dimensions.

We evaluated how \sysname{} impacts designers' authoring experience with 13 expert designers in a controlled in-lab user study.
Participants found \sysname{} provided a more intuitive, fluid, and engaging color authoring experience and facilitated their theme exploration and visual continuity management, a notable improvement over existing tools. 
Additionally, findings suggest that \sysname{} could seamlessly be integrated with Adobe After Effects, potentially shifting the color planning to post-animation stages and reducing reliance on initial color decisions in Adobe Illustrator. 
This research thus has threefold contributions:

\begin{itemize}
    \item A \textbf{formative study} with 6 expert motion graphic designers that identify the typical MG creation workflow, understand common approaches employed for MG color design and pinpoint challenges in color authoring.
    \item A \textbf{prototype system, \sysname{}}, co-designed with experts, to support \emph{context-aware}, \emph{group-wise}, and \emph{multi-level} color authoring tailored to MG videos.
    \item An \textbf{expert evaluation} confirms the effectiveness of \sysname{}, revealing its advantages and limitations. 
    The derived insights shed light on future investigation of video authoring designs.
\end{itemize}

\section{Related Work}

% \jian{The related work is very tool-oriented, missing a review of the relevant empirical studies on color (palettes). Your Sec 2.3 seems to talk a little about it. Sec 2.2 of InfoColorizer may help you on this. Since you present the needs of the formative study first (also as your first contribution), you may consdier to move Sec 2.3 to the front.}

\subsection{Color Design and Interactive Color Authoring Tools}
Color design, which consists of choosing specific colors and allocating them to distinct design elements, is crucial in designers' workflow, greatly shaping the visual aesthetic appeal \cite{eiseman2000pantone}. 
Instead of examining colors in isolation, 
designers typically ``feel'' colors within a composition as a whole \cite{kim2017thoughts} because the relationships among colors can profoundly change the overall aesthetic and emotional resonance \cite{whitfield1990color}. 
% As a result, considering colors within the context of the entire composition is important for informing design decisions \cite{}. 
To craft balanced and captivating color compositions, designers frequently turn to predefined color palettes for inspiration. 
Popular platforms like Adobe Color\footnote{\url{https://color.adobe.com/create/color-wheel}} and ColorBrewer\footnote{\url{https://colorbrewer2.org}} offer palette suggestions rooted in established harmonic principles \cite{wyszecki2000color}.

\subsubsection{Color Authoring Tools for Static Designs}
\rev{
Commercial tools like Adobe Photoshop~\footnote{\url{https://helpx.adobe.com/ca/photoshop/using/color-adjustments.html}} and Illustrator~\footnote{\url{https://helpx.adobe.com/ca/illustrator/using/adjusting-colors.html}} offer advanced color adjustment features for static designs, tailored to rasterized images and vector graphics, respectively. 
While these tools are powerful, they often present a steep learning curve, especially for novice designers, and require much user manual effort. 
To address this, the research community has focused on developing more intuitive and user-friendly color adjustment tools.}
% Innovative tools have been developed for color authoring in images, largely enhancing traditional color picker interfaces \cite{olsen2007evaluating}. 
For instance, Color Portraits \cite{jalal2015color} pioneered the \emph{in-context} exploration, enabling users to play with color interrelationships (\ie, size, position) using color swatches. 
However, the palette exploration and its assignment to the intended design remain separate. Bridging this gap, tools like Color Triads \cite{shugrina2020nonlinear} and Histomages \cite{chevalier2012histomages} introduced a triangular-shaped palette and color histogram, respectively, to allow users to directly recolor images by manipulating the palette. 
Specifically tailored for fine-grained color blending, Playful Palette \cite{shugrina2017playful} employs a free-form unstructured palette for artwork colorization.
% Generally, these solutions can be categorized as palette-based image editing.
% They are mainly designed for hue and saturation adjustments while neglecting lightness which is typically adjusted via curve-based methods.
\rev{Extending the palette-based editing,}
a recent work, ColorfulCurves \cite{chao2023colorfulcurves}, unifies the palette and the curve-based solution to allow for interactive regional lightness adjustment for images.
\rev{
Beyond supporting color manipulation, 
recent research has expanded into data-driven approaches to recommend colors for various types of graphic design, such as infographics \cite{yuan2021infocolorizer}, advertising landing pages \cite{qiu2022intelligent}, and image-centric designs \cite{shi2023stijl}. These methods aim to optimize aesthetic appeal and enhance the design's effectiveness in conveying the intended message.

Although current research offers promising color authoring methods for static graphic designs, they cannot be directly applied to motion graphics. 
This is primarily due to the dynamic nature of motion graphics, where there is a need to consider consistent color transitions, layer interactions, and the evolving visual narrative. Therefore, more tailored approaches are needed to address the unique challenges posed by the temporal complexity of motion graphics.
}
% \rev{Tailored to vector graphics, Adobe Illustrator offers various color adjustment tools, \eg, individual color shift~\footnote{\url{https://helpx.adobe.com/ca/illustrator/using/adjusting-colors.html}}, color group adjustments~\footnote{\url{https://helpx.adobe.com/ca/illustrator/using/color-groups-harmonies.html}}, and automatic recoloring~\footnote{\url{https://helpx.adobe.com/ca/illustrator/using/recolor-artwork.html}}. }

% While existing tools provide promising solutions for static design color exploration, there remains a gap in tailored solutions for motion graphics video color authoring. 
% The closest approach is proposed by Du~\etal~\cite{du2021video}, which employs a spatial-temporal geometric palette for video recoloring. 
% However, this method is designed for pixel-based, real-world captured videos and is anchored to the image color space. 
% In contrast, this paper focuses on motion graphic videos crafted from vector graphics.
\rev{
\subsubsection{Color Authoring Tools for Videos.}
Existing research mainly focuses on color exploration for static designs, with limited attention to video color authoring. 
Du~\etal~\cite{du2021video} proposed a spatial-temporal geometric palette for video recoloring but limited to pixel-based, real-world videos that are constrained by image color space. 
In contrast, our paper focuses on vector graphic-based motion graphic videos, predominantly edited using Adobe After Effects.
However, color modifications are challenging in After Effects due to its limited color adjustment features. 
A few color management plugins for After Effects are developed to compensate for such deficiencies. 
Color Shift~\footnote{\url{https://aescripts.com/color-shift/}} enables color range remapping but lacks precision and flexibility. 
Ray Dynamic Color~\footnote{\url{https://aescripts.com/ray-dynamic-color/}} provides more precise color adjustments across scenes by linking colors with the same value in different layers to a unified color palette. 
However, its simplistic palette for color manipulation uses a basic, non-ordered linear square design, lacking structured organization and failing to provide detailed information about color proportions and timeline distribution.
% This limitation makes assessing color composition and navigating color links challenging, particularly in complex motion graphics with a wide range of colors. \jian{what do you mean by navigating color links?}
Consequently, this limitation makes it hard to effectively assess overall color composition and to track or manage the connections between colors, especially in complex motion graphics with a wide range of colors. 
While ColorSwap~\footnote{\url{https://aescripts.com/colorswap/}} allows users to search and replace individual colors in the MG video, users cannot simultaneously manipulate multiple colors while maintaining some properties. 

Our paper presents an interactive color palette design %that enhances existing plugin capabilities 
with greater flexibility, accuracy, and scalability. 
Our approach provides group-wise color editing in motion graphics, a feature that is strongly demanded but not available in existing methods. 
Such a feature could ease the cognitive burden and reduce the manual effort for designers working on complex motion graphic projects.
}
% Inspired by the benefits of interactive palettes in color exploration for static design coloring tools, our aim is to extend such a palette-driven approach in motion graphic video color authoring.

% \subsection{Intelligent Tools for Motion Graphics Authoring}
\subsection{Intelligent Tools for Motion Authoring}
Motion graphics are pieces of animated graphic design that use text, graphical elements, and animation to convey information in a concise and immersive way \cite{amini2015understanding}. 
Over time, studies have investigated to understand this complex domain, producing a set of design principles \cite{thomas1995illusion} that govern the effective use of motion and animation in communication.
% The creation of motion graphics often presents significant challenges. 
\rev{The creation and authoring of motion often present significant challenges. }
Dominant commercial tools like Adobe AfterEffects offer advanced capabilities but come with a steep learning curve that can be unfriendly for novices and even some intermediate users. 
Recognizing the need for ease, platforms like Animaker\footnote{\url{https://www.animaker.com} \label{Animaker}}, Vimeo\footnote{\url{https://vimeo.com} \label{Vimeo}}, and Doodly\footnote{\url{https://www.voomly.com/doodly}} offer template-based tools, making motion graphic creation accessible for beginners. 
However, the template-based solution has limited flexibility, and customization is not straightforward.

To seek the balance between simplicity and flexibility, several innovative solutions have emerged. 
For instance, Katika \cite{jahanlou2022katika} provided an end-to-end system, enabling amateurs to produce animations through a three-fold approach: generating shots from scripts, sourcing animations from a crowd-contributed library, and modifying animations using modular motion bundles. 
Similarly, Kitty \cite{kazi2014kitty} introduced a sketch-based platform, which empowers artists to craft animated, interactive visuals by delineating entities and their interrelationships. 
This is further enriched by an underlying graph model that visually conveys dynamic interactions and causalities as viewers engage with the content. 
Complementing these, Ma~\etal~\cite{ma2022layered} proposed a layered authoring tool for 3D animation authoring, employing a timeline sequencer for easy preset stylization effects, while also offering a node-graph-based UI for further customizations. 

Despite significant advancements in motion authoring, there remains a notable gap in tools for color authoring. 
Current trends typically separate motion authoring from element design, leading to the challenges of making seamless color adjustments. 
In this paper, \rev{we present the integration of a video preview to enhance contextual understanding for users as they adjust colors, and further design and develop a color authoring tool tailored for motion graphics.}
% In this paper, we propose seamlessly integrating a video preview to provide contextual information while users modify colors, and further investigate how to provide a more intuitive and efficient authoring experience for post-animation color adjustments.
% \jian{a bit vague on ``further investigate how to ...''}

% \subsection{AI-assisted Motion Graphics Generation and Augmentation}
\subsection{AI-assisted Motion Graphics Generation}
Creating motion graphics is complex and difficult, which typically involves narration writing, script planning, coordination of visuals, and animation implementation \cite{jahanlou2020challenges}. 
% To streamline the fragmented and tedious nature of this process, researchers have employed AI techniques, primarily making their efforts to automatically generate motion graphics and augment existing videos with animated visuals.
To streamline the fragmented and tedious nature of this process, researchers have employed AI techniques, primarily making their efforts to automatically generate motion graphics.
% \textbf{Automatic MG video generation.}
Narration-driven methods dominate the field of automatic motion graphics (MG) generation. 
Key works include AutoClips \cite{shi2021autoclips}, which automated data video creation through a two-step process of clip library usage and optimal clip arrangement. 
DataParticles \cite{cao2023dataparticles} simplified animated unit visualization creation using block-based and language-oriented authoring, obviating the need for expertise in tools like D3.js and AfterEffects. 
Similarly, Data Player \cite{shen2023data} and Wonderflow \cite{wang2023wonderflow} presented a narration-driven data video generation approach but with an enhanced focus on narrative-animation synchronization.
Broadening the scope of data-centric motion graphics, TakeToons \cite{subramonyam2018taketoons} presented a script-driven approach for performance animation where authors annotated scripts with animation events. 
These annotated scripts would then be compiled into a story model, bridging gaps in the fragmented production process. 
Doc2Video \cite{chi2022synthesis}, introduced a pipeline applied for general transforming documents to talking head motion graphic video solution, consisting of document parsing, script planning, and video synthesizing, greatly automating the content transformation process.

% \textbf{Augmenting existing videos using animated visuals.}
% Enhancing videos with animated visuals has grown in popularity, predominantly aiming to enhance communication and viewer engagement.
% Notably, Visual Captions \cite{liu2023visual} enhanced video conferencing with a visual intent model to predict and display the most apt visuals during conversations.
% RealityTalk \cite{liao2022realitytalk} addressed live presentation tools' limitations, offering real-time speech-driven graphics control. 
% Similarly, Chironomia \cite{hall2022augmented} took this further in remote presentations, overlaying interactive visualizations on webcam footage, augmenting the presenter's gestures, which not only allows presenters to control visual elements but also enhances their expressiveness.
% Different from the presentation-centric efforts, Sporthesia \cite{chen2022sporthesia} ventured into sports video augmentation by detecting textual insights, converting them into animations, and seamlessly embedding them, enriching the viewing experience.

These works offer promising solutions for automatically generating motion graphics. 
Yet, a significant gap remains: none of them carefully consider the color choice, despite its influence on viewer perception and emotional resonance \cite{von1970theory}. 
In this field, human designers invest significant time in fine-tuning color choices. 
Recognizing this gap, our paper investigates color authoring in motion graphics.
We believe that addressing this aspect will bring such generative solutions closer to real-world applications, ensuring enhanced visual aesthetics.

\section{Formative Study}
\label{sec:formative_study}
A formative study was conducted to gather insights from expert motion graphic designers, with the aim of 
understanding: (1) what are the typical workflows of motion graphics video creation; (2) how designers approach early-stage color exploration and late-stage color adjustments; and (3) what are the challenges they encounter during color authoring. 

\subsection{Participants and Procedure}
We conducted semi-structured interviews with six domain experts, all of whom have received professional training in motion graphics design, hold a formal degree in Design, Arts, or Digital Media Arts, and have over four years of experience in crafting explainer MG videos. 
% Each expert has contributed to the creation of more than \x such videos. 
Further demographic details are provided in Table~\ref{tab:interview_demographic}.
In the subsequent explanations, we refer to our experts as E\#.

\begin{table}[tb]
\caption{The table records our interviewees' demographic information, including genders, range of ages, occupations, and experiences of motion graphics video creation in years (denoted as Exp. in the table).}
\vspace{-3mm}
\label{tab:interview_demographic}
\begin{tabular}{lcclc}
\toprule
\textbf{ID} & \textbf{Gender} & \textbf{Age}   & \textbf{Occupation}  & \textbf{Exp.} \\
\midrule
E1 & F & 25-29 & Motion Graphics Designer       & 7                       \\
E2 & F & 25-29 & Visual Designer                & 6                         \\
E3 & M  & 20-24 & Student in Digital Media Arts & 4                         \\
E4 & F & 25-29 & Motion Graphics Designer       & 5                         \\
E5 & M & 25-29 & Graphics Designer              & 6                          \\
E6 & F & 20-24 & Motion Graphics Designer       & 4                          \\
\bottomrule
\end{tabular}
\end{table}

The interviews began with background questions about participants' professional roles and experiences in MG video creation. 
Participants were then asked to introduce their typical workflow for designing and creating explainer MG videos by showing us intermediate results of one recent work and demonstrating their process in their preferred editing tool. 
Subsequent discussions revolved around the early exploration and late-stage adjustments of color choices in the MG video, and the common challenges they encountered in color authoring.
Each interview lasted approximately one hour, during which both audio and screen-sharing were recorded.

\subsection{Data Analysis}
\label{sec:data_analysis}
We conducted the thematic analysis of the interview data and built an affinity diagram to investigate the patterns of designers' workflow and themes of challenges they encountered. 
Two researchers independently analyzed and open-coded the transcribed interviewees' responses.
Then, they discussed and reconciled any discrepancies in the coding process to ensure a consistent and accurate representation of the participants' perspectives.
Subsequently, the researchers used affinity diagramming to categorize the initial codes onto cards.
Through an iterative discussion and organization of the codes, we identified several recurring patterns and themes from the collected data.

\subsection{Key Findings}
Our findings regarding the workflows (Section~\ref{sec:workflow}), approaches to handle color exploration and adjustments (Section~\ref{sec:color_explore}), and challenges in color authoring (Section~\ref{sec:challenges})have been summarized as follows. 

\subsubsection{Workflows for MG Videos Creation}
\label{sec:workflow}

\begin{figure*}[htbp]
    \centering
    \includegraphics[width=.98\linewidth]{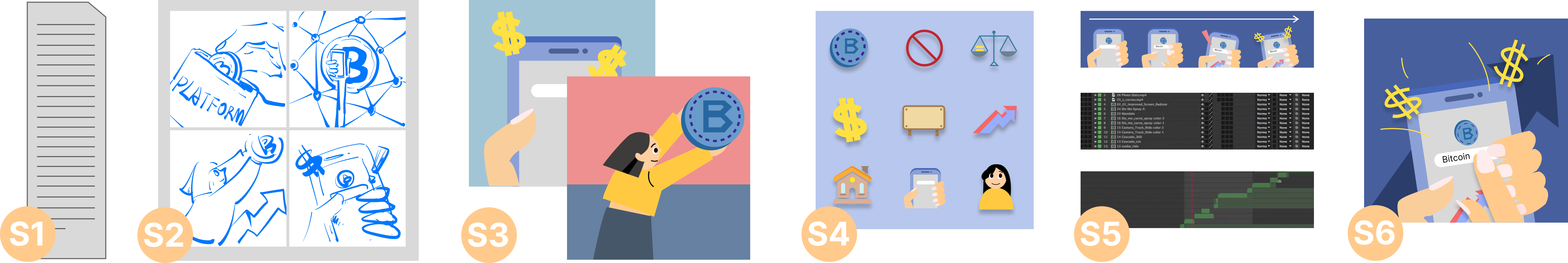}
    \vspace{-3mm}
    \caption{An example provided by E4 explaining Bitcoin, showing the typical workflow to create an explainer motion graphics video. The original narration is missing, thus we replaced it with a representational figure for S1. We have received consent from E4 to include this in the paper.}
    \label{fig:workflow}
    \Description{This figure presents a detailed view of the typical workflow in creating an explainer motion graphics video. It is organized in a sequence of steps (ordered from left to right), each representing a stage in the video production process: S1 is a long textual doc icon used to represent the narration of the video; S2 contains four line-drawing images that outline the main scenes for the video; S3 contains two colored images which are two keyframes in the video; S4 shows a collection of vector shapes made by the designer; S5 shows two screenshots of the animation implementation in the Adobe After Effects and the resulted animated shot; and S6 is a poster frame of the video indicating the final compiled outcome video.}
\end{figure*}

\begin{table*}[htbp]
\caption{Six stages of motion graphics video creation.}
\small
\vspace{-3mm}
\label{tab:workflow}
% \scalebox{1.0}{
\begin{tabular}{lll}
\toprule
\textbf{Stage} & \textbf{Description}   \\
\midrule
S1. Narrative Analysis &   
Analyzing client-provided scripts to extract logic and segment into scenes based on story flow.         \\
S2. Storyboard Sketching &   
Drafting line drawings of key scenes to outline the video progression and crucial transitions.              \\
S3. Styleframe Design  &     
Developing 3-5 keyframes (depending on scenes and budget) to set the style and define theme colors.      \\
S4. Element Creation &    Creating visual elements with tools like Adobe Illustrator, finalizing detailed color choices.       \\
S5. Scene Animation       &   Animating scenes individually using After Effects to maintain organization and reduce complexity.                   \\
S6. Video Compilation      &   Assembling animated segments using Premiere Pro or AE to produce a cohesive MG video.                 \\
\bottomrule
\end{tabular}
% }
\end{table*}

Crafting an explainer motion graphics (MG) video is a complex, time-consuming, and full trial-and-error process.
This process involves a multi-step workflow commonly adopted by designers, consisting of six major steps: narrative analysis, storyboard sketching, styleframe design, element creation, scene animation, and video compilation.
An example of intermediate results from these steps, provided by E4, is illustrated in Figure~\ref{fig:workflow}. 
Although communications with clients and discussions within the design team primarily take place during the static design stages (S1 to S4), the process is not always linear. 
As experts emphasized, \pqt{when doing animations (S5), we often need to go back to earlier stages to make adjustments.}{E2}, a loop that includes communications with clients and iterative refinements.

% \begin{itemize}[leftmargin=3em]
%     \item[\textit{Step 1.}] \textit{Narrative analysis.} 
%     Clients often provide a narration script, occasionally accompanied by an audio track. 
%     Designers will then analyze this narrative, extract its logic, and segment it into scenes based on the story flow.
%     \item[\textit{Step 2.}] \textit{Storyboard sketching.} 
%     During this stage, designers draft line drawings of key scenes, showing essential elements and rough motions without colors, to establish the video's progression and highlight important transitions.
%     \item[\textit{Step 3.}] \textit{Styleframe design.}
%     After the storyboard, designers create style frames by selecting 3-5 keyframes, determined by the number of scenes and project budget. 
%     These fully colored frames establish the video's aesthetic tone and define the overall theme colors.
%     \item[\textit{Step 4.}] \textit{Element creation.}
%     Next, designers craft the required visual elements using tools such as Adobe Illustrator. 
%     Most detailed color decisions are finalized in this phase.
%     \item[\textit{Step 5.}] \textit{Scene animation.}
%     Designers then animate using After Effects, tackling scenes individually. 
%     This modular method reduces the complexity and keeps the workspace organized.
%     \item[\textit{Step 6.}] \textit{Video compilation.}
%     Finally, these animated segments are assembled in Premiere Pro or AE, resulting in a unified explainer MG video.
% \end{itemize}

\subsubsection{Early Exploration and Late-stage Adjustments of Colors. }
\label{sec:color_explore}

Colors are always a crucial part in MG video creation. 
Prior research \cite{shugrina2020nonlinear} revealed that color design is an iterative process in static graphic designs. 
Our findings dive into the color design process in MG videos, particularly, how designers explore the color choices in an early stage and how they make necessary adjustments later.

\textbf{How do designers explore color choices in an early stage?}
The initial discussions about colors between designers and clients happened during the styleframe design stage (S3). 
Initially, designers will select a palette of 3-5 colors, guided by the emotional tone of the story, brand impressions, and the target audience. 
Then, designers apply these color choices to several keyframes, \ie, styleframes. 
They demonstrate the interplay of these colors within the frames, offering a glimpse of their potential impact in the final video. 
It helps clients form a concrete picture of the final video's visual aesthetics and styles.
During this stage, designers may present alternative color schemes for clients to consider, while also welcoming client suggestions for different color combinations. 
It facilitates early color exploration, allowing for quick and cost-effective adjustments.
Ideally, this stage is supposed to clearly set the client's expectations regarding the video's style and firmly establish the color themes to avoid later changes. 

\textbf{Why do color adjustments often happen in late stages?}
Despite the conversations in the styleframe design stage defined the styles and color themes, late-stage (\ie, post-animation) color adjustments can not always be avoided.
As noted by E1, \qt{while we aim for a linear and smooth process, there's always that back-and-forth on tones that can drive us crazy.} 
This primarily results from the differences between the designers' envisioned outcome and the client's expectations, a gap often noticeable when comparing static styleframes with the final video.
All experts interviewed had encountered such late-stage color adjustment requests, which generally fall into two categories: tweaks to a specific color tone (\eg, shifting from one shade of blue to another) and changing the entire color scheme (\eg, revising theme colors from green and yellow to blue and pink).
For the former, E6 shared us her experience, \qt{in one project, the client gave us the wrong blue tone by mistake, a color that used a lot, here and there in the video. They spotted this when all animations were already done. This forces us into a lot of painful changes}.
E4 provided us with her experience on the later types of adjustment requests, \qt{Once, we produced a 150-video series for history education, each about 30 seconds long. The client wanted a unified feel, so we applied a consistent color theme across the series. Just as we neared completion, the client jumped in, asking, `Can we make the visuals more colorful?' You can imagine how such a request is annoying.}

\textbf{How do designers handle late-stage color adjustments?}
Such late-stage color adjustments are very tedious and challenging, E5 called it \qt{a disaster and tough to rescue}.
In such cases, designers will \pqt{go back to create new static styleframes, and re-evaluate the potential impact of the color changes on the final video.}{E2}.
Once the new color choices are confirmed by clients, designers will \qt{revisit the original Illustrator element files to modify colors of related elements} and \pqt{avoid directly make adjustments on AE source files.}{E5}.
However, \pqt{if the existing files are cluttered and messy, it sometimes becomes more practical to craft new elements from scratch, replacing the previous ones entirely.}{E3}
This highlights the importance of early-stage project planning including \pqt{file naming, graphic grouping, and layer organization}{E5}; however, \pqt{the chaos is the normal}{E4}.
Experts all noted that disorganized project planning and collaborative work, which are often the cases in their work, make using original files for post-animation color adjustments extremely challenging.

\subsubsection{Challenges in MG Video Color Authoring}
\label{sec:challenges}
Given that late-stage color adjustments in MG video creation is frequent but challenging, we further discuss with experts about the specific challenges and frustrations they encountered. 

\textbf{C1: Absence of immediate contextual feedback.}
MG video color authoring is notably different from static graphic design because MG videos incorporate dynamic, moving elements. %, whereas static graphic designs are unchanging. 
The color choices in MG videos not only influence a single frame but also affect the perception of sequences of frames.
This distinction makes it important for designers to ensure that edited color effects cohesively align with the story's narrative scaffolds, maintaining consistent visual harmony throughout the video.
E3 emphasized that \qt{color choices should `act' with the story rather than standing as isolated elements.} 
Yet, as E2 expressed, the current process often feels like \pqt{assembling a puzzle without knowing what the final image looks like. Each piece feels right, but then it doesn’t fit into the larger scheme of things.}
This common frustration arises from two reasons: the fragmented workflow and the lack of real-time previews.
E3 expressed frustration about the frequent need to \qt{switch between Illustrator and After Effects, particularly for minor tweaks} which \qt{can be very annoying and disrupt the holistic design experience.}
The separation of element coloring from its resulting effect, based on its interaction with other elements and storyline flow, causes this back-and-forth.
Designers thus expressed their wish for \pqt{live previews of changes, so we don’t work blind.}{E1}. 

\textbf{C2: Inability to group colors for joint manipulation.}
For static designs, designers appreciate the efficiency of jointly adjusting hue and saturation for multiple regions, like using Photoshop's hue/saturation adjustment layer\footnote{\url{https://helpx.adobe.com/photoshop-elements/using/adjusting-color-saturation-hue-vibrance.html}}. 
It is ideal to afford rapid experimentation with color tones.
However, this convenience is missing in MG video authoring, as suggested by E4, \qt{it would be super useful to have that where we can tweak colors for elements spread across different scenes in MG videos}.
E2 explained the inherent reason leading to this challenge: \qt{each element is crafted separately to keep our files neat and organized}. As a result, each element is isolated, making modifications on different colors across scenes difficult. 
This mandates designers to \pqt{modify the smallest unit each time.}{E2}.
Alternatively, E5 mentioned that \qt{there do exist ways to globally adjust, like adding some universal filters [to the whole video], but let's be real, there is no magic fix. Proper [color] adjustment is complex, it is shifting some colors here in this way, then fine-tuning some colors in that way, there is no one-size-fits-all deal.}
In summary, existing tools only support the two extreme solutions, either tweaking every detail or just blindly applying a universal filter to the entire video, with no room in between for grouping target colors flexibly for strategic adjustments.
Such limitation makes it tedious and time-consuming to maintain or change a consistent color theme across videos, particularly for extensive projects.
% E3 shared her experience on this: \qt{Once, we produced a 150-video series for history education, each about 30 seconds long. The client wanted a unified feel, so we applied a consistent color theme across the series. Just as we neared completion, the client jumped in, asking, ``Can we make the visuals a bit more vibrant?'' 
% You can imagine how much workload that such a simple request needs.} 
This highlights the necessity for supporting joint manipulation for grouped colors.

\textbf{C3: Difficulty in progressively authoring colors.}
An MG video, much like a static image, can contain a lot of colors. 
Yet, it is the dominant colors that leave a lasting impression, revealed by visual saliency theory \cite{gopalakrishnan2009salient} and also confirmed by our interviewees. 
Designers naturally adopt a top-down strategy: \pqt{starting from the predominant colors that cover large areas or appear frequently, then on secondary hues, and finally refining less obvious details, such as character hair.}{E4}
The benefit of this strategy is noted by E1, \qt{prioritization makes me more clear to the larger picture I want to paint.}
However, the current tools challenge this approach. Designers must \pqt{repeatedly play the video, find dominant colors, adjust them, and move to the next,}{E2} such hierarchy should be kept on track mentally. 
E3 complained that \qt{as the number of elements grows and the video gets longer, keeping track [such hierarchy] mentally becomes hard.}
She attributed this to the manipulation nature adopted by current tools, calling it \qt{object-by-object disconnected logic}.

\subsection{Design Goals}
Driven by the observed challenges in MG video color authoring and our discussions with motion graphics professionals, we derived the following design goals, each addressing one specific challenge in MG video color authoring, which guided the development of \sysname{}. 
% \jian{is there a 1:1 mapping between C's and D's? If not, cite the C's in the following D's.}

\textbf{D1: Integrate color editing with video playback for real-time contextual feedback.}
The temporal nature of MG videos requires a continuous and cohesive narrative. 
By merging color editing with video playback, designers can immediately see how color changes impact the storyline. 
This not only speeds up the design process but also ensures that the narrative flow remains undisturbed by color modifications.

\textbf{D2: Enable both individual and group-wise color adjustments.}
In MG video authoring, designers have dual needs: refining individual elements and harmonizing colors across multiple elements for style consistency. 
This balance stems from two distinct design perspectives: the \emph{object-driven} focus on isolated components, and the \emph{scene-driven} approach, which emphasizes holistic visual cohesion. 
To support both, it is crucial to offer different views that highlight individual objects and the overall color palette within scenes. 
By facilitating both types of adjustments, we simplify the editing process, enabling designers to swiftly transition between detailed tweaks and broad thematic edits.

\textbf{D3: Present an overview of color distribution with adjustable detail levels.}
Understanding the color distribution in a video is important for informed color design decisions and prioritizing designers' edits. 
Our study indicates that color dominance in an MG video hinges on its visual presence and recurrence across frames. 
Therefore, it is essential to ensure these aspects are reflected in the color distribution overview. 
In addition, an adjustable overview is necessary to allow designers to zoom in on specifics or zoom out for a broader perspective, giving them flexibility in how they approach and execute their color adjustments.

\section{\sysname{}}
\begin{figure*}[htbp]
    \centering
    \includegraphics[width=\linewidth]{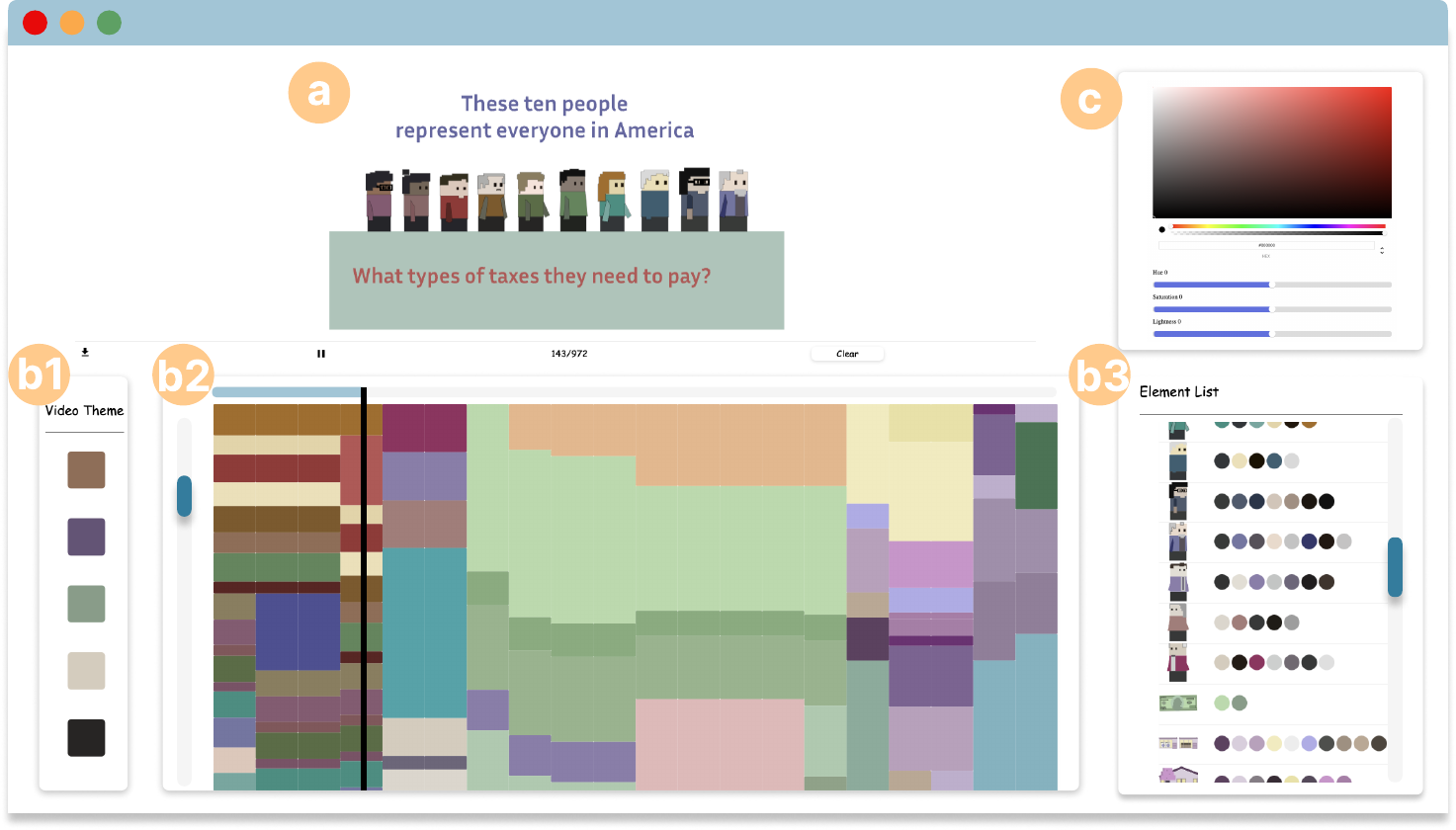}
    \vspace{-6mm}
    \caption{User Interface of \sysname{}, consisting of (a) \textit{Video Playback}, (b1) \textit{Video Theme} summarizing the dominant colors throughout the video, (b2) \textit{Scene Palette} illustrating detailed color distribution on the timeline, (b3) \textit{Element List} detailing colors for every element, and (c) \textit{Color Picker}. These views are both interactive and synchronized.}
    \label{fig:interface}
    \Description{This figure illustrates the user interface of \sysname{}, a tool with multiple interactive and synchronized components designed for video color management. The interface is divided into five parts. Part (a), positioned in the middle top of the interface, is the \textit{Video Playback} area and this is where the video is displayed, allowing users to view the content they are working on. Part (b), located in the left bottom corner, is labeled as the \textit{Video Theme} view, showing five colored rectangles which are the extracted dominant colors in the video. Part (c), situated at the middle bottom, is the \textit{Scene Palette} describing all the colors used in the video, along with their respective proportions, mapped out across the video timeline. This allows for a detailed understanding of color distribution throughout the video. Part (d), in the right bottom area of the interface, is the \textit{Element List} listing each element within the video, accompanied by the specific colors associated with each of these elements.  Part (e), at the right top corner, comprises an RGB color selector alongside three sliders for adjusting hue, saturation, and lightness.
    }
\end{figure*}

\sysname{} is a color authoring tool tailored to motion graphics videos by providing different interactive color representation views summarizing colors from broad video overviews down to detailed scene and element perspectives. 
% \jian{do we really need the stuff below for a conference paper? the design goals can be easily referred to like `(D1)' after the subsection title. I would merge 4.1 and 4.2 to name it `System Overview'}
% Below, we detail \sysname{}’s interface (Section~\ref{sec:interface}) and characterize its functionalities from three aspects: 1) how color information is extracted from an MG video (Section~\ref{sec:color_info_extraction});
% 2) how different color representation views are constructed (Section~\ref{sec:color_view_construction}); and 3) what interactions can be performed with \sysname{} (Section~\x - Section~\x).
% The functionalities aim to tackle each challenge and design goal identified in Section~\ref{sec:formative_study}.
% We add direct references for clarity.
% The detailed implementations are described in Section~\ref{sec:implementation}, followed by a simulated usage scenario to walk through \sysname{} (Section~\ref{sec:usage_scenario}).

\subsection{System Overview}
\label{sec:interface}
\sysname{} has three main panels: the Video Playback (Figure~\ref{fig:interface}a), the Color Representation (Figure~\ref{fig:interface}b), and the Color Picker (Figure~\ref{fig:interface}c). 
Within the Color Representation panel, there are three views: 1) the Video Theme, summarizing the dominant colors and their proportions throughout the video; 2) the Scene Palette, illustrating detailed color distribution on the timeline; and 3) the Element List, detailing colors for every element. 
These views are both interactive and synchronized: modifications in one view automatically update the others.
They are presented at all times, side-by-side as the layout shown in Figure~\ref{fig:interface}.

% \subsection{Extracting Colors from MG Video}
% \label{sec:color_info_extraction}
We require MG video imported to \sysname{} must be in \emph{Lottie} format\textsuperscript{\ref{lottie}} a JSON-based animation file format that can be exported from current most widely used motion graphics creation tools and platforms such as Adobe AfterEffects, Figma, and Canva. 
\rev{
The choice of the \emph{Lottie} format over Adobe After Effects (\emph{.aep}) for development is driven by its compatibility across web platforms, smaller file sizes, and efficient web rendering performance. 
\emph{Lottie}'s open-source framework provides sufficient accessibility and benefits community collaboration.
Its JSON-based structure simplifies programmatic implementation. 
Therefore, we utilize \emph{Lottie} as our testbed. 
However, it's important to note that our proposed designs are feasible with the \emph{.aep} format, though their implementation might require more effort due to its complex structure.
}
While Lottie handles both vector (SVG) and raster (image) elements, we only consider MG videos constructed from SVG elements due to the notable difference in colorization between images and SVGs.  
In Lottie, each SVG shape is represented as a layer, characterized by static Bezier paths and animated property transformations such as how its color, opacity, scale, position, and rotation change across keyframes. 
From Lottie, we extract each SVG element's color information including \textit{RGB value}, \textit{opacity}, \textit{approximated size}, and \textit{timing}. 
The RGB and opacity are directly accessible, whereas a color's size is derived from its shape's Bezier path bounding box area. 
These properties align with keyframes, marking the start and end time of each color's appearance.

\subsection{Constructing Color Representation Views}
\label{sec:color_view_construction}

\subsubsection{Video Theme View.}
The process of extracting theme colors is based on the frequency and proportion of the color appearing throughout the video. Initially, all colors present in the video are converted into \rev{CIE LAB} representation. Subsequently, the K-Means algorithm is applied to derive $k$ clusters of colors, simply based on the DeltaE distance. Within each cluster, the color with the highest proportion is selected as the theme color for that cluster. In our system, we have chosen $k$ to be 5.

\subsubsection{Scene Palette View.}
The scene palette aims to represent the color distribution throughout the timeline of a video, essentially illustrating how the colors and their size change over time (D3). In other words, the scene palette serves as a visual representation of the color scheme in a video over time. There are two primary considerations for displaying this visualization in a way that effectively communicates the color distribution to users. Firstly, there is the challenge of encoding the colors from the video into color rectangle units within the palette. Secondly, there is the matter of how these different color rectangles are arranged or stacked within the palette.

\textbf{Visual encoding.}
The height of each colored unit in the palette is determined by its corresponding shape in the video. However, this involves converting visual perception from 2D to 1D (\ie, encoded from a 2D area to a single height dimension). To address this, we have adjusted the height of a colored rectangle in the palette to be proportional to the square root of the bounding box size of the respective shape in the video. Specifically, the height is calculated using the formula: 
${Height_c} = \alpha \times \sqrt{{Shape_c}}$,
where $\alpha$ is a modifiable parameter used to control the zoom level of the color palette.
% Another problem need to consider is different scenes can have unbalanced number of colors, thus the visualization of colors
% The time step is flexible to handle the scalability issue when the video is complex, long, and the number of scenes is large. 
% $$ h = xxx $$

\textbf{Color merging.}
Grouping identical colors offers a more streamlined and comprehensive view of the color distribution for viewers. 
For instance, in the given video example, there are numerous instances of cash. 
It can be beneficial to merge these instances to view and edit the color collectively.
Thus, we merged all identical colors, even though they might appear on different elements. 
The rationale is that the scene palette aims to provide an overview of how colors are distributed throughout the timeline. 
Moreover, according to the \rev{Gestalt law of similarity}\cite{wertheimer1938gestalt}, humans perceive identical colors as a group. 

\rev{
\textbf{Color sorting.}
We organize colors in the CIE LAB color space based on their DeltaE distances from $(0,0,0)$.
This sorting strategy forces similar colors to be close together, which aims to reflect designers' common practice of using slightly varied colors to maintain visual continuity without monotony. 
Therefore, the scene palette can better align with human perceptual tendencies and offer a more intuitive grasp of color distribution.
The sorting is executed only at the initial loading phase, and any subsequent user modifications to colors do not trigger a re-sorting. 
This decision is made to reduce users' confusion and cognitive load because re-sorting post-modification could disorient users by changing the positions of colors they have already interacted with, making it hard to track both modified and unmodified colors.
However, our method has a limitation. 
Within a scene, colors are relatively stable, but as frames progress and new elements with different colors are introduced, the consistency of a color's position across frames is not guaranteed. 
For instance, as shown in Figure~\ref{fig:linked_view}, the introduction of the orange color can shift the starting vertical position of the existing green color. 
This can result in a color's position varying across different frames, which is a challenge that needs a more sophisticated sorting algorithm.}
% \xinyu{I am not sure if this addresses R3's comments of ``more details are needed on how the top-to-bottom ordering of colored is performed for the frames. There seems to be consistent ordering across frames, but how is this accomplished, and what are the design challenges there?'', please help to revise.}
% \jian{looks good to me}

\subsubsection{Element List View.}
In the element view, each component can comprise various sub-shapes, like a person having eyes and a hat. 
This approach complements the scene palette view. 
In the Scene Palette, colors form the palette, disregarding their association with specific elements, and focusing on the overall image. 
Yet, designers need to maintain precise control over each detail, facilitating fine-tuning at later stages. 
The element-centric perspective of this view enables swift navigation to targeted elements for refinements, making the design process more efficient and precise.

\begin{figure*}[htbp]
    \centering
    \includegraphics[width=\linewidth]{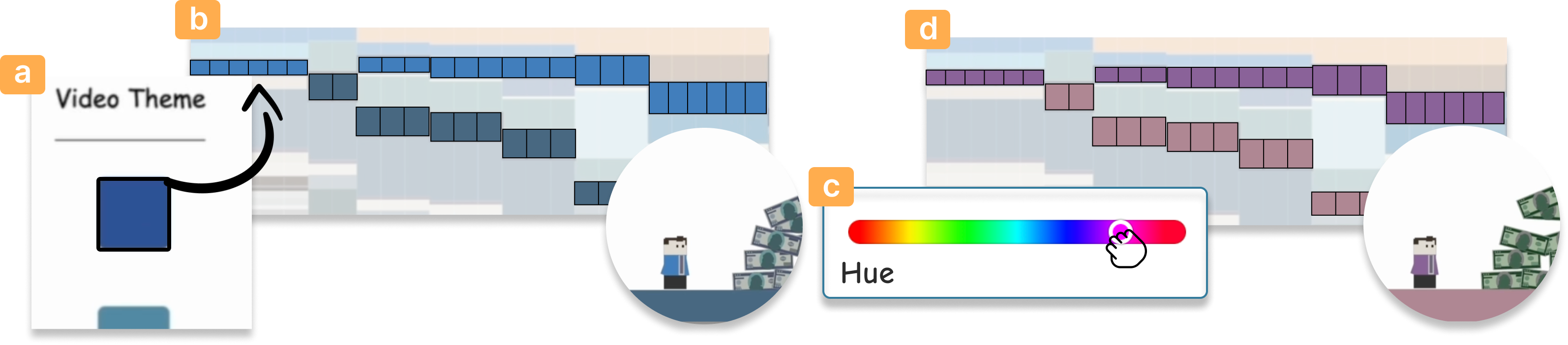}
    \vspace{-6mm}
    \caption{\sysname{} features group-wise color adjustments. Users can follow these steps: (a) choose a theme color in the \textit{Video Theme} view, (b) the \textit{Scene Palette} will automatically select colors similar to the chosen theme, (c) manipulate the HSL (Hue, Saturation, Lightness) slider, (d) the selected group of colors will be changed together.}
    \label{fig:group_wise}
    \Description{This figure illustrates the steps of how users can make group-wise color adjustments with the automatic selection mechanism. From left to right, there are four steps: (a) choosing a theme color in the Video Theme view, (b) the Scene Palette will automatically select colors similar to the chosen theme and there is a circle image showing the corresponding frame, (c) manipulating the HSL (Hue, Saturation, Lightness) slider, (d) the selected group of colors will be changed together and a there is a circle image at the corner showing the resulted frame after the colors changed.}
\end{figure*}

\begin{figure*}[htbp]
    \centering
    \includegraphics[width=\linewidth]{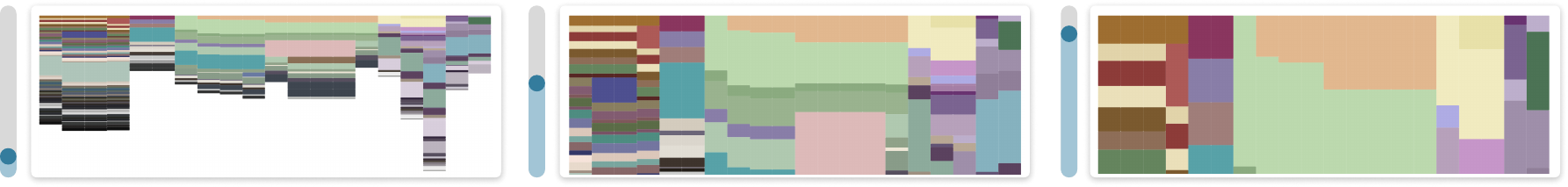}
    \vspace{-6mm}
    \caption{\sysname{} provides users with the flexibility to adjust the level of details in the \textit{Scene Palette} to handle minor color blocks and encourage the progressive workflow.}
    \label{fig:level_of_detail}
    \Description{This figure shows how the \textit{Scene Palette} alters its appearance when the zoom-in slider is adjusted. The figure consists of three images aligned horizontally, demonstrating a progression from left to right. The first image on the left depicts a low zoom-in level where the overview of the color distribution across the video timeline is presented but the details are not highly distinct. The middle image shows the moderate level of zoom-in where a clearer view of the color distribution is visible, revealing more details compared to the first image. The third image, on the right, displays a high zoom-in level where the palette shows a detailed and close-up view, making even the minor details and subtle variations in color distribution distinctly visible.}
\end{figure*}

\subsection{Interacting with Scene Palette}

\subsubsection{Linking Color to Elements in the Video.}
\sysname{} optimizes the authoring experience by allowing the designers to modify colors while staying within their authentic
video frame context. 
Hovering over a color block in the scene palette automatically shifts the time marker to the corresponding frame, presenting it in the video playback area. 
Simultaneously, the associated SVG element pulses with a black animated border, offering designers a clear visual cue about the element in focus and its relationship with surrounding elements. 
Any color changes will be reflected instantly, with the video dynamically re-rendering in real-time, ensuring designers have an immediate, contextual feedback loop.

\subsubsection{Modifying An Individual Color.}
Within the scene palette view, users can modify a specific color by clicking its corresponding color block.
\rev{The color block may span multiple frames, indicating the continuous presence or intermittent appearance of a certain color across different frames.}
The targeted block will be highlighted with a black animated outline, signifying its selection. 
While users click on one specific frame to select a color, by default, this selection is applied to this color across all frames (\ie, the color block), given that elements typically maintain consistent color throughout a video.
Once the color is selected, users have access to an RGB color picker and HSL sliders for precise adjustments.
\rev{However, if designers opt to alter an element's color in specific frames, a double-click on a single color frame can isolate this change to only that frame. 
This creates a smooth color transition animation effect, where the color transits to the specified value for that frame and then returns to the original color.
The animation function is handled by \emph{Lottie}.
}

\subsubsection{Modifying Grouped Colors.}
\label{sec:group-wise}

\sysname{} features the group-wise color modifications, shown in Figure ~\ref{fig:group_wise}.
\rev{
Users have two options to group colors: automatic and manual. 
In auto-group, selecting a color in the Video Theme view groups similar colors from the Scene Palette based on a preset similarity range. 
Users can further refine their selection by adding overlooked colors or removing unwanted ones in the Scene Palette. 
Alternatively, manual-group allows users to select individual colors in the Scene Palette to create a group. 
}
% Users can select multiple color blocks from the scene palette by clicking on them sequentially. 
Once colors are grouped, they will be highlighted with a black outline and made editable. 
The RGB color picker is deactivated during this phase. 
Users can modify the color properties using sliders for hue, saturation, or lightness in the HSL space, adjusting one property at a time. 
Our design choice encourages HSL adjustments for grouped colors to ensure harmonious shifts by making their hues (or saturations or lightness) change uniformly. 
Editing in the RGB color space to make all selected colors identical is not allowed. 
This decision stems from the feedback from the pilot studies with our early \sysname{} prototype that users rarely desired this RGB interaction for multiple colors representing diverse elements. 
Adjusting distinct colors to a single RGB value could unintentionally homogenize visual elements. 
While designers sometimes aim for a sense of cohesion among different elements, they often opt for varied gradients rather than identical colors.

\rev{
To manage out-of-range cases in our group-wise color adjustments in the HSL space, we apply specific rules for saturation, lightness, and hue. 
If saturation or lightness adjustments for any color exceed their respective bounds, we clamp these values to their maximum or minimum limits. 
For hue, we implement a circular adjustment: values exceeding 360 degrees loop back to the start, while negative values loop from the end, allowing for smooth transitions. Any out-of-range behavior in one color does not impact the others in the group. 
In \sysname{}, all color adjustments are instantly visible in both the scene palette and the video, allowing users to intuitively detect out-of-range issues by noticing when certain colors stop changing. 
Thus, we currently do not offer explicit feedback for out-of-range adjustments.
}

\subsubsection{Adjusting Focus and Interest Levels.}

To adapt to designers' progressive authoring workflows and accommodate varying design needs and detail intricacies, the scene palette incorporates both zooming and scrolling capabilities.
\textit{Zoom Functionality:}
As shown in Figure~\ref{fig:level_of_detail}, the zoom slider, ranging from 0\% to 100\%, is tailored for a progressive color authoring workflow and ensures the smooth manipulation of tiny color blocks. 
Initially, the scene palette is scaled to mid-size so that dominant colors are clearly visible.
As mentioned earlier, colors within the palette are strategically sorted by LAB color representation, the relatively large granularity of color adjustment helps in maintaining a coherent and harmonious color scheme throughout the video.
As users progressively delve deeper into the palette, smaller color blocks can prove challenging to interact with. 
Here, the zoom slider becomes useful. 
By sliding it upward, the scene palette enlarges, expanding all color blocks proportionately.
This allows designers to magnify and engage with even the minutest color segments comfortably.

\textit{Scrolling Mechanism:} 
After zooming, if a designer's color of interest isn't centrally placed, a simple vertical scroll can reposition it for easier manipulation. 
Moreover, recognizing scenarios with an extensive number of scenes or elongated timelines, a horizontal scroll option ensures users can traverse the scenes on their timeline.

By offering users the flexibility to magnify, minimize, and navigate their color spectrum, \sysname{} ensures designers have optimal control and visibility at every stage of their progressive color authoring process.

\subsection{Inspecting Visual Continuity}

\begin{figure}[htbp]
    \centering
    \includegraphics[width=\linewidth]{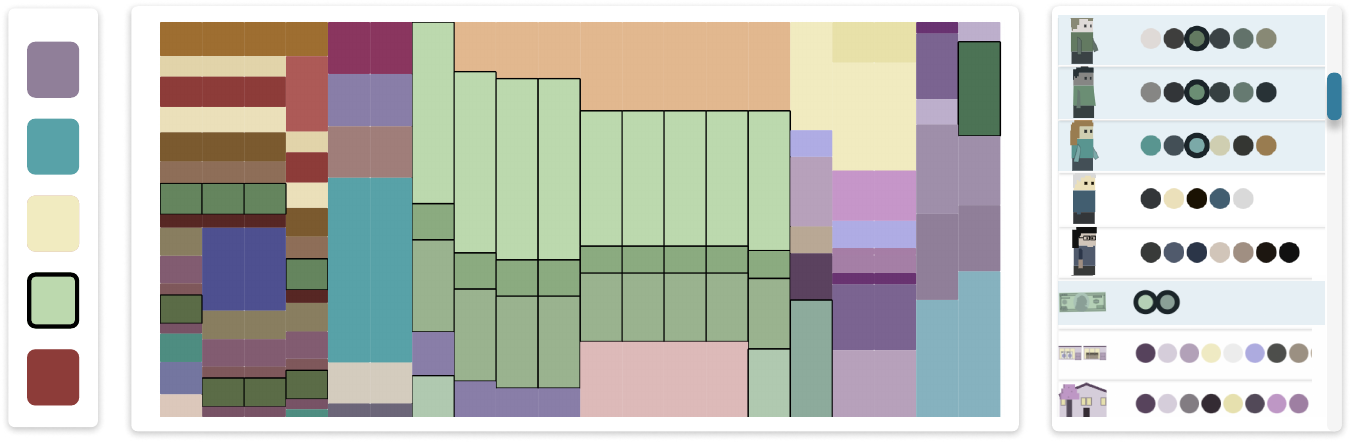}
    \vspace{-6mm}
    \caption{The three color representation views, \ie, Video Theme, Scene Palette, and Element List are synchronized to help designers inspect how theme colors are distributed throughout the timeline and interconnected scenes, highlighting with the black outline.}
    \label{fig:linked_view}
    \Description{This figure highlights the synchronization of three key color representation views within \sysname{}: the Video Theme, Scene Palette, and Element List. The black outline is employed to highlight areas of focus or interaction across these views. When users click on one color in a certain view, relevant colors in other views will be highlighted with the black outline.}
\end{figure}

\begin{figure*}[htbp]
    \centering
    \includegraphics[width=\linewidth]{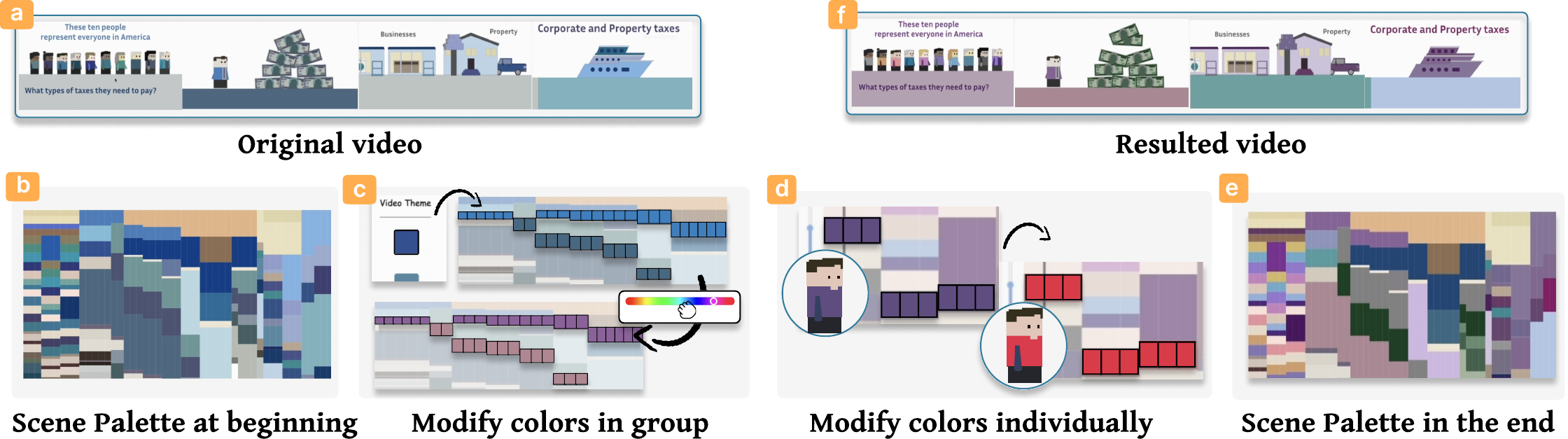}
    \vspace{-6mm}
    \caption{The process of a user modifying the MG video color with \sysname{} where (a) shows the original video and (b) presents the current \textit{Scene Palette}. Users can (c) modify the overall color theme through group-wise manipulation and (d) refine individual colors until satisfied. (e) is the appearance of the \textit{Scene Palette} after the modifications and (f) shows the resulted video.}
    \label{fig:usage_scenario}
    \Description{This figure illustrates the step-by-step process a user follows to modify the color of a motion graphics video using \sysname{}. The figure is divided into several key parts, each depicting a different stage of the color modification process: part (a) displays the original video whose theme colors are blue and grey; part (b) shows the current \textit{Scene Palette} of the original video; part (c) demonstrates how users can modify the overall color theme of the video. This is done through group-wise manipulation, allowing for broad changes to the video's color scheme; part (d) illustrates the process of refining individual colors with an example of turning the clothes from purple to red; part (e) reveals the appearance of the \textit{Scene Palette} after these modifications have been applied; part (f) presents the resultant video following the color modifications. }
\end{figure*}

Visual continuity is important for achieving cohesive storytelling in videos. 
In line with this, we've designed the three color representation views (\ie, Video Theme, Scene Palette, and Element List) to be synchronized.
We introduce an intuitive interaction mechanism to help designers inspect how theme colors are distributed throughout the timeline and interconnected scenes. 

As shown in Figure~\ref{fig:linked_view}, when designers hover over a specific theme color in the Video Theme view, it triggers a visual cascade. 
The chosen color, along with its visually similar counterparts, will be spotlighted in the Scene Palette view using an animated black outline. 
This visual similarity is determined by measuring the color distance in the Lab space, with a predefined threshold.
Parallel to the Scene Palette highlight, elements bearing these colors within the Element List are simultaneously highlighted. 
It ensures designers see the thematic interplay across all views in real-time.
% If users modify one theme color, all highlighted colors can be automatically adjusted by shifting the hues, yet meticulously maintaining their respective gradients. 
% This operation is similar to performing the group-wise color adjustments in the scene palette.
By providing this immediate visualization of theme colors and their varied similar colors throughout the timeline, designers can intuitively gauge the video's visual continuity.

\subsection{Implementation Details}
\label{sec:implementation}
All showcased motion graphics (MG) videos in this paper were created using the Figma platform and later exported to the Lottie format, with the version of \emph{LottieFiles Figma v41}. 
It should be noted that this version captures color data slightly different from the official documentation on the Lottie website. Despite this, all essential color details can be extracted from both the Figma version and the official version, with minor implementation adjustments. 
The MG video rendering is implemented based on Lottie.js.
The front-end was developed using the React framework.
Moreover, the D3.js library was utilized in constructing the scene palette.

\subsection{System Walkthrough}
\label{sec:usage_scenario}
% \jian{a usage scenario should be described in present tense.}

We explain how designers use \sysname{} by walking through a typical usage scenario workflow, \rev{as shown in Figure~\ref{fig:usage_scenario}}.
Imagine Stella is a seasoned motion graphics designer, now working on creating an explainer MG video about American taxation \rev{(Figure~\ref{fig:usage_scenario}a)}. 
After crafting each segment to explain each type of tax people need to pay, she presented the initial draft to her client, an educational institution. 
The initial color palette was a blend of blue, black, and gray tones. 
However, the client envisioned the project evolving into a vibrant educational series catering more to a younger audience. 
Stella realized she needed to make adjustments to colors to make them more vibrant and engaging for children.

With the renewed vision, Stella initiated the recoloring process. 
First, she exported the MG video into the Lottie format. 
Stella then launched \sysname{} and imported the Lottie JSON file.
The video is quickly displayed in the Video Playback panel. 
\sysname{} identified the existing theme colors and displayed them with their proportions in the Video Theme view.
The detailed color distribution was shown on the Scene Palette \rev{(Figure~\ref{fig:usage_scenario}b)}, with all SVG elements with their colors represented as colored bubbles listed in the Element List.

Stella considered the clients' feedback carefully, and some cheerful colors that would resonate with young learners came to her mind: purple, green, and orange tones. 
She hovered over the current dominant colors on the Video Theme.
As her cursor glided over the dark blue, the Scene Palette responded dynamically, highlighting areas near the blue where adjustments would take place.
She clicked on the Scene Palette, activating the editing mode. 
Her artistic intuition guided her to move on to the hue slider, morphing the blue into a rich purple \rev{(Figure~\ref{fig:usage_scenario}c)}. 
She repeated the process, changing the reserved light blue to a lively pink. 
By changing the theme colors, the overall video looks more resonant with children now, but Stella knew that the transformation was not complete.

She sought to enhance the secondary, yet frequently appearing color, grey, to harmonize with the new theme. 
% Her focus turned to the notable appearances of red within the video. 
She selected the grey blocks in the Scene Palette, switching them to a softer pink that melded perfectly with the new palette.
% She then noticed a green segment that needed adjustment, but only in one specific scene. 
% With a quick double-click, she isolated the scene, changing the green to a complementary yellow without affecting the other portions of the video.
As she went deeper into the intricacies of the color palette, Stella realized that some elements required a more detailed touch. 
The clothing of the characters, though minor in proportion, played a vital role in the aesthetics. 
Aware of the minute adjustments needed, she utilized the zoom function on the Scene Palette, making the smaller blocks more accessible for fine-tuning \rev{(Figure~\ref{fig:usage_scenario}d)}.
To improve the project, Stella navigated to the Element List panel for an even closer inspection. 
Here, she swiftly examined each element, ensuring even the smallest details harmonized with the vibrant new theme and aligned with the semantic meaning of the element.

Stella pressed the Play button to watch the video from start to finish. 
Witnessing the video becomes colorful and aligns with youthful vibrancy \rev{(Figure~\ref{fig:usage_scenario}f)}, she feels satisfied.
Confident that her client would appreciate the polished colors, she exported the newly adorned Lottie file, eager to show it to her client.

\section{Expert Evaluation}
We conducted a user study to evaluate the authoring experience of \sysname.
In particular, we aim to answer the following research questions:
1) How participants use \sysname{} and if there are any authoring strategies they employ in \sysname{}?
2) How well \sysname{} can facilitate color authoring in MG video?
3) How might \sysname{} impact designers’ workflow in color authoring for MG videos?

\subsection{Participants}

\begin{table}[tb]
\caption{The table records our user study participants' demographic information, including genders, range of ages, occupations, and experiences of motion graphics video creation in years (denoted as Exp. in the table).}
\vspace{-3mm}
\label{tab:evaluation_demographic}
\begin{tabular}{lcclc}
\toprule
\textbf{ID} & \textbf{Gender} & \textbf{Age}   & \textbf{Occupation}  & \textbf{Exp.} \\
\midrule
P1 & F & 20-24 & Graphic Design Student & 5 \\
P2 & F & 25-29 & Graphic Designer & 4 \\
P3 & F & 25-29 & Graphic Design Student & 3 \\
P4 & F & 20-24 & Graphic Designer & 4 \\
P5 & F & 25-29 & Motion Graphic Designer & 5 \\
P6 & F & 25-29 & Motion Graphic Designer & 4 \\
P7 & F & 25-29 & Motion Artist & 5 \\
P8 & F & 25-29 & Motion Artist & 4 \\
P9 & M & 25-29 & Interaction Designer & 7 \\
P10 & F & 20-24 & Motion Graphic Designer & 3 \\
P11 & F & 20-24 & Motion Graphic Designer & 3 \\
P12 & F & 25-29 & Graphic Design Student  & 5 \\
P13 & F & 20-24 & Motion Graphic Designer & 3 \\
\bottomrule
\end{tabular}
\end{table}

We recruited 13 experts in the domain of motion graphics creation (\ie, 12 females, 1 male) through email lists and personal connections. 
All participants hold a formal design degree and have more than 3 years of experience in MG video creation. 
The detailed demographic information is listed in Table~\ref{tab:evaluation_demographic}.
The studies were conducted remotely via Zoom. 
Participants accessed \sysname{} through a web browser. 
They received a \$25 CAD Amazon gift card for the 60-minute session.

\subsection{Task and Design}
We designed two tasks to evaluate how \sysname{} assists designers on MG video color authoring regarding multiple aspects of creativity and system usability. 
Both tasks require users to tweak colors in MG videos from a provided MG video to simulate the post-animation color modification scenarios. 
We made the MG video adapted from Vox\footnote{\url{https://www.vox.com/videos/2019/12/20/21028676/tax-poor-rich-data-video}} explaining the tax mechanism in America. 
Given that our tasks only focus on colors, and users do not need to create elements and implement animations by themselves; thus, we set the time limit to be relatively short: 10 minutes for each task.
Tasks are designed to be from a more flexible open-ended scenario to a more constrained closed-form scenario. 
Details are as follows.

\textbf{Task 1} focuses on encouraging participants to freely explore and play with \sysname{}. 
Participants were asked to refine the colors to ensure an effective narrative transition and visual harmony throughout the video. The usage of colors fully depends on the designers' personal taste without any constraints.

\textbf{Task 2} aims to simulate the cases of changing the color theme from one to another.
Responding to client feedback, they want to integrate the video into an existing educational series, thus designers need to align the theme colors to the series. 
This entails modifying the current color scheme to match the specified theme colors, ensuring a cohesive look and feel across the series. 
The specified theme colors are navy blue (\#023E73), bright blue (\#085CA6), dusty rose red (\#8C4265), and pale gold (\#D9B97E).

\subsection{Procedure}
After signing the consent form, participants were given an introduction to the study procedure, duration, data collection,
and other important information. 
Participants were then provided with a brief training session on \sysname{}, during which an example was used to demonstrate how to modify colors using the tool. 
Participants were allowed to play with the tools using a pre-determined video example independent of the user task video. 
Once they became comfortable using the tools, participants were introduced to the above tasks sequentially. 
During both task sessions, participants were asked to ``think aloud.''
After completing the two tasks, participants were asked to
rate their experiences with \sysname{} using a post-study questionnaire, which included the System Usability Scale (SUS) \cite{brooke1996sus} rating on Likert scales. 
To gain a deeper understanding, semi-structured interviews were conducted to collect their feedback on \sysname{}. 
All study sessions were audio and video recorded.

% \begin{figure}[htbp]
%     \centering
%         \includegraphics[width=\textwidth]{figures/study3.pdf}
%         % \label{fig:123123}
%     \vspace{-8mm}
%     \caption{Results of Creativity Support Index (CSI) regarding six factors: exploration, enjoyment, expressiveness, collaboration, immersion, and results worth effort.}
%     \label{fig:csi_results}
% \end{figure}

\subsection{Quantitative Results}
% \jian{don't forget to come back to your design goals and research questions in the results}

% \subsubsection{System Usability Scale (SUS)}
% \begin{table*}[t]
% \caption{Question statements of System Usability Scale (SUS).}
% \vspace{-4mm}
% \begin{tabular}{ll}
% \toprule
% ID  & Question & ID  & Question  \\
% \midrule
% Q1  & I think that I would like to use this system frequently.  \\
% Q2  & I thought the system was easy to use. \\
% Q3  & I found the various functions in this system were well integrated. \\
% Q4  & I would imagine that most people would learn to use this system very quickly. \\
% Q5  & I felt very confident using the system.  \\
% \midrule
% Q6  & I found the system unnecessarily complex. \\
% Q7  & I think that I would need the support of a technical person to be able to use this system. \\
% Q8  & I thought there was too much inconsistency in this system. \\
% Q9  & I found the system very cumbersome to use. \\
% Q10 & I needed to learn a lot of things before I could get going with this system. \\
% \bottomrule
% \end{tabular}
% \end{table*}

\begin{table*}[t]
\caption{Question statements of System Usability Scale (SUS).}
\vspace{-4mm}
\scalebox{0.8}{
\begin{tabular}{llll}
\toprule
\textbf{ID}  & \textbf{Question} & \textbf{ID}  & \textbf{Question}  \\
\midrule
Q1  & I think that I would like to use this system frequently. & Q6  & I found the system unnecessarily complex. \\
Q2  & I thought the system was easy to use. & Q7  & I would need the support of a technical person to be able to use this system. \\
Q3  & I found the various functions in this system were well integrated. & Q8  & I thought there was too much inconsistency in this system. \\
Q4  & I would imagine most people would learn to use this system very quickly. & Q9  & I found the system very cumbersome to use. \\
Q5  & I felt very confident using the system.  & Q10 & I needed to learn a lot of things before I could get going with this system. \\
\bottomrule
\end{tabular}}
\end{table*}

\begin{figure*}[htbp]
    \centering
    \begin{subfigure}[]{\textwidth}
        \includegraphics[width=\textwidth]{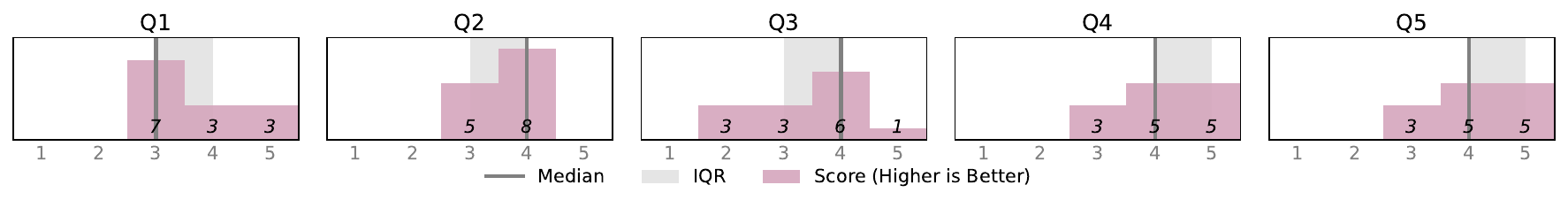}
        \label{fig:figure1}
        \vspace{-0.5cm}
    \end{subfigure}
    % \vspace{-0.5cm}
    %\vspace{-10mm}
    \begin{subfigure}[]{\textwidth}
        \includegraphics[width=\textwidth]{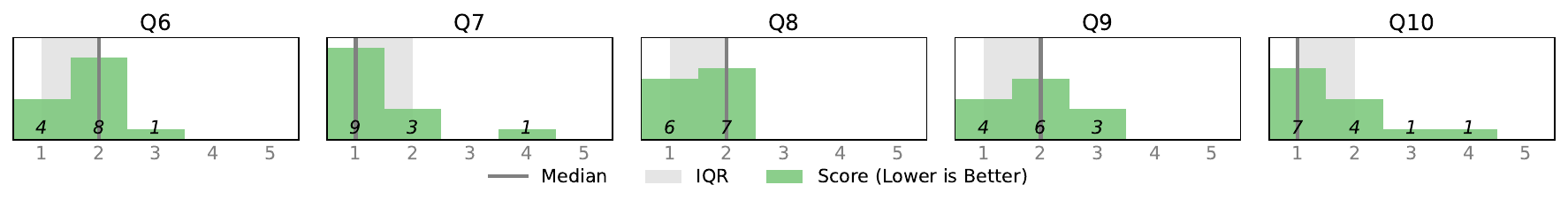}
        \label{fig:figure2}
    \end{subfigure}
    \vspace{-9mm}
    \caption{Results of participants rated System Usability Scale (SUS). For Q1 to Q5, the higher the better; for Q6 to Q10, the lower the better.}
    \label{fig:sus_results}
    \Description{This figure presents the results of a System Usability Scale (SUS) assessment, where participants rated various aspects of usability. The results are organized into two groups based on the nature of the questions: for questions Q1 to Q5, a higher rating indicates better usability while for questions Q6 to Q10, a lower rating is preferable.}
\end{figure*}

Participants rate their perceived system usability on a 5-point Likert scale (1-strongly disagree, 5-strongly agree).
The results, shown in Figure~\ref{fig:sus_results} illustrated a generally positive response to the \sysname{} tool, indicating that users found the tool to be user-friendly and efficient in facilitating color authoring for MG videos.
They rated that they were able to learn how to use the system quickly and modify the colors easily, felt confident to use the system, and would like to use the system frequently.

% \subsubsection{Creativity Support Index (CSI)}
% Participants rated six creativity support factors with scores from 0 (worst) to 10 (best). 
% Figure~\ref{fig:csi_results} shows the individual CSI score for each factor, \ie, enjoyment, exploration, expressiveness, immersion, collaboration, and results-worth-effort. 
% Overall, participants felt \sysname{} provided great creativity support.
% Specifically, participants found it easy to explore different color possibilities in \sysname{}, fun to use the system, felt engaged during the usage, and satisfied with the produced results using the \sysname{}.

\subsection{Usage Pattern Analysis}
\label{sec:usage_pattern}
\begin{figure}[htbp]
    \centering
    \includegraphics[width=\linewidth]{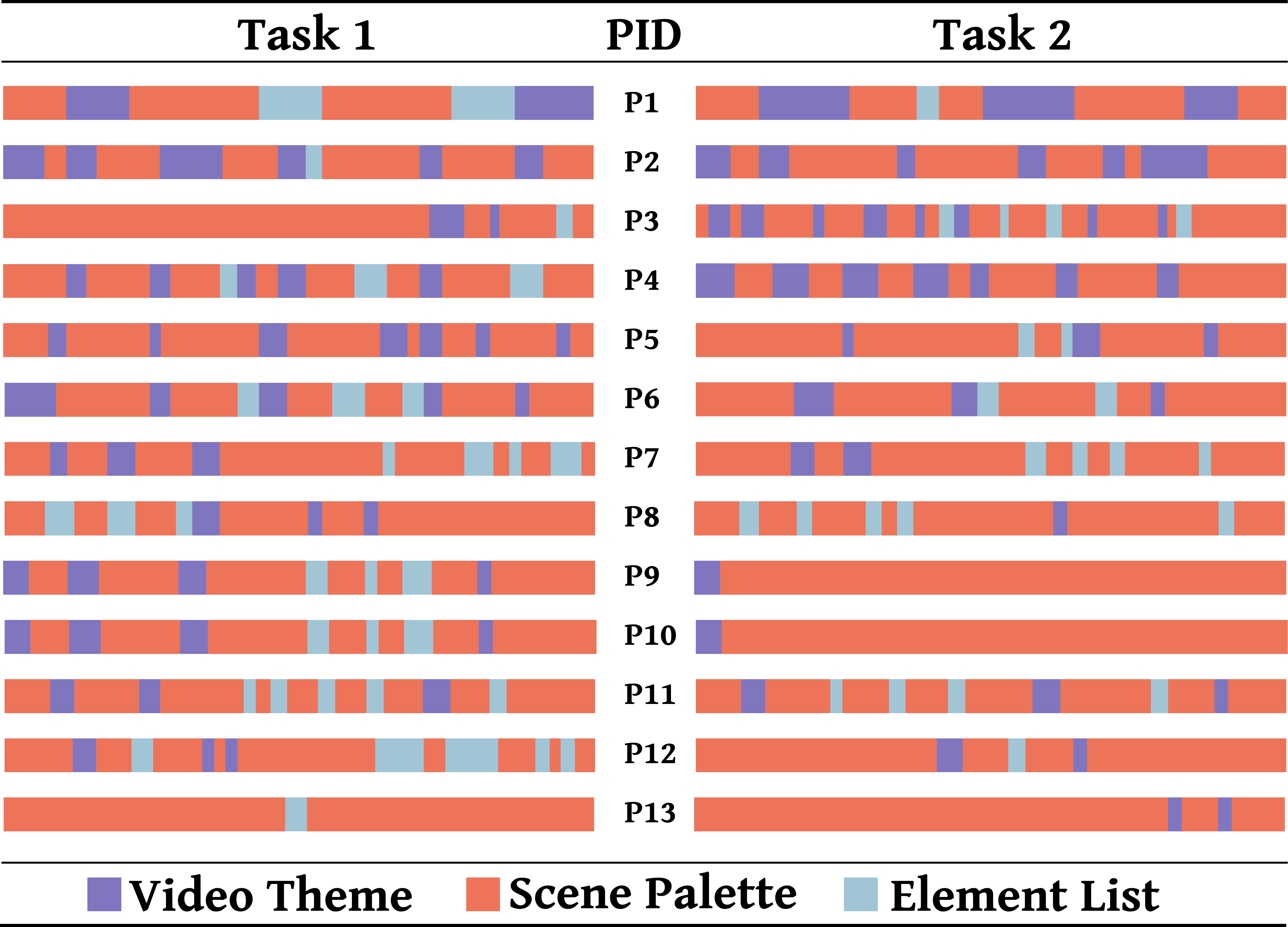}
    \vspace{-6mm}
    \caption{Visualizing the usage pattern of how users interact with the three color representation views in \sysname{} by manually coding the screen recordings.}
    \label{fig:usage_pattern}
    \Description{This figure provides a visual analysis of how users interact with the three main color representation views in \sysname{}: the \textit{Video Theme}, \textit{Scene Palette}, and \textit{Element List}. The data for this analysis was gathered by manually coding user interactions from screen recordings during their use of the system. Detailed analysis of usage pattern is presented in Section~\ref{sec:usage_pattern}.}
\end{figure}

We systematically analyzed the interaction patterns of designers with \sysname{} by manually coding the screen recordings. 
This process helped to understand how the participants navigated through the three provided color representation views: Video Theme, Scene Palette, and Element List over the allotted task duration. 
All participants used the full 10 minutes on each task.
The interaction pattern is depicted in Figure~\ref{fig:usage_pattern}, where the x-axis is the entire duration of each task.
We identified the following strategies adopted by participants.

\textbf{Blending usage of the three color representation views.}
Overall, participants predominantly utilized Scene Palette as a base, while also transitioning to the Video Theme and Element List views to gather detailed insights and make specific adjustments. 
This pronounced preference for the Scene Palette hints at its alignment with the intrinsic authoring requirements of designers, satisfying the need for an overview of color distribution across the timeline (D3).
Moreover, the utilization of all three views, despite being used in varying degrees, facilitates a flexible and dynamic approach to color manipulation, where designers alternate between object-driven and color distribution-driven strategies. 
Hence, the multi-level functionalities of the Scene Palette, Video Theme, and Element List prove to be instrumental in accommodating the diverse authoring demands at different stages of the design process.

\textbf{Non-linear yet Progressive Engagement with Different Views.}
Participants exhibited a non-linear yet progressive engagement with the different views offered by the tool, predominantly anchoring their tasks on the Scene Palette and fluidly navigating between the other panels as needed. 
Generally, users tended to explore the Video Theme, not for direct color modifications but as a navigational aid, exploiting its linkage with the Scene Palette to group and adjust colors harmoniously with the chosen theme. 
As they progressed, a gradual shift towards the Element List was observed, particularly for detailed, nuanced adjustments. Nevertheless, this pattern was not universal, with instances like P8 starting with the Element List while using the Video Theme panel later.

\textbf{Task-Oriented Panel Usage.}
A notable difference in panel engagement regarding the Video Theme and Element List was observed between Task 1 and Task 2. 
In Task 1, characterized by its open-ended structure, participants tend to lean more towards the Element List, utilizing it extensively to fine-tune individual elements. 
This contrasts with the behavior noted during Task 2, where the focus shifted predominantly to the Video Theme to enact broader, thematic modifications. 
Interestingly, a subset of participants (P9, P10, P13) completely bypassed the Element List in Task 2, signifying a distinct strategy that aligns with the task's demand for cohesive changes. 
This analysis underscores the adaptable usage patterns of the tool, where users selectively leverage different panels in accordance with the specific demands and nature of the task at hand, optimizing for both detail-oriented and macro thematic alterations.

\subsection{Qualitative Results}

By adopting the data analysis approach as described in Section~\ref{sec:data_analysis}, we derived the following key findings. 

\subsubsection{Perceived Benefits of Scene Palette in Enhancing Color Understanding and Authoring}
The Interactive Scene Palette is a key feature in \sysname{}, serving two main functions: it helps designers visually track color distribution across the timeline and makes it easier to change colors directly in the video, without adjusting each element separately. 
Users pointed out several benefits of the Scene Palette, noting that it improves their experience with video color understanding and authoring.

\textbf{Intuitiveness.}
Participants found the Scene Palette to be both intuitive and effective. 
Almost every participant's first comment for \sysname{} is \qt{the [scene] palette is very intuitive.}
It offers a comprehensive view of the color scheme, allowing them to quickly grasp the proportions and dominance of different colors in different scenes. 
One participant noted, \qt{This nicely displayed color overview provides me with a clear sense of which colors stand out and how they harmonize with each other.}

\textbf{Focus.}
The Scene Palette has been highly praised for maintaining focus throughout the color-authoring process.
Many Participants (9/12) mentioned that this feature considerably reduces the distraction often caused by the need to jump between individual elements.
As pointed out by P7, \qt{for me, if I go back to Illustrator to adjust a color, in most cases, I'd adjust the element's shapes and lines as well because the switching is costing so I will do all potential changes at once. But this is also interrupting my thoughts around colors. But with this tool (\sysname{}) is different. It's like having a clear path without any interruptions and distractions pulling me away, keeping me very focused on feeling and playing with colors.}

\textbf{Playfulness.}
A notable highlight from the feedback was the enjoyment and engagement the Scene Palette introduced. 
P5 provided an interesting analogy of the authoring experience: \qt{it is like I am customizing a LEGO character.}
Other participants mentioned terms like \qt{interesting} and \qt{cool} to describe their interaction with the tool. 
One user enthusiastically commented, \pqt{You can really play with it, experimenting with colors becomes a joyful task rather than a chore.}{P9}.

\subsubsection{Group-wise Color Manipulation Enabled Rapid Theme Exploration and Visual Continuity Management}
During the study, we observed participants heavily rely on the Scene Palette to modify colors in groups.
Such group-wise color manipulation involved diverse elements from various frames, which is not feasible in existing platforms like AI and AE, is \rev{one} unique benefit brought by \sysname{}.
Participants described it as \qt{the most impressive and useful feature}, which largely improved the efficiency of color modifications, especially theme exploration and visual continuity management.
P11 shared that \qt{this (group-wise modification) makes changing themes feel so free, so simple. Otherwise, switching from one theme to another would be a massive headache, no one wants to try out different theme possibilities.}
Another participant emphasized its importance in maintaining the overall mood and theme consistency, stating, \pqt{It's more about setting the direction, the theme, and the overall feeling, which becomes much more manageable with this feature.}{P4}.

Overall, the group-wise color manipulation in \sysname{} was seen as a significant leap forward, offering a way to visually plan and adjust color distributions across timelines, which was previously unachievable. 
This innovative approach not only saves time but also unlocks the potential for greater creativity and freedom in theme exploration.

\subsubsection{Multi-level Color Representation Views Facilitated Progressive Color Authoring Process}
The multi-level views design of \sysname{} was appreciated by participants for its structured approach to the color authoring process. 
P12 commented, \qt{The three panels align well with my work style, guiding me naturally from large color blocks to theme colors, and finally to detailed refinements using the Element List.}
Participants found this structured design particularly helpful in tasks with defined color theme changes, such as Task 2, where it provided a clear modification path, helping them focus and avoid distractions. 
\pqt{In tasks with specific changes, the system clearly guides where and how to make adjustments}{P6}.
Even in tasks without clear guidelines like Task 1, the multi-level representation served as a valuable starting point, inspiring an organized approach to color alterations. 
P11 mentioned, \qt{The hierarchical design gave me a clue on where to start, encouraging a structured approach.}
Overall, \sysname{} promotes a top-down, structured approach to color authoring, making the process more organized, especially for tasks with defined color modifications.

\subsubsection{Potentials for Shifting Color Planning to Late Stages}
\sysname{} focuses on the color authoring for MG videos, especially during the stages after animation where colors need to be adjusted. 
We wanted to know how designers might use \sysname{} in their current workflows for MG video creation and how it might change things for them.
All participants thought \sysname{} would work great as a plugin in Adobe After Effects. 
They expressed that \pqt{I really liked how the tool showed colors on the timeline, which fits nicely with how things are done in After Effects.}{P7}.

With \sysname{} working as a plugin in After Effects, a large portion of participants felt they preferred to shift most of the color work to when they finished the animation.
P3 shared, \qt{if had it (\sysname{}) in AE, I'd use AI (Adobe Illustrator) less. I will just make elements in it (Illustrator) without worrying about colors at first, maybe just using some grays. After finishing the animation, I'd play with colors using this tool.} 
One participant even being more ambitious, \pqt{I feel for some simple ones (MG videos), we even do not really necessarily need Illustrator anymore, given AE itself has some basic graphical element creation functionalities and your tool handles colors so well.}{P5}.
This opinion was quite popular among participants (10/12), with many planning to spend less time on color choices in Illustrator, opting for fast color choices or just grays at first. 
P8 explained to us the reason for such color planning shifting, \qt{When you spend a lot of time choosing colors in Illustrator, sometimes they don't feel the same in the final video. The way colors change from one scene to the next is very important in videos, and it's hard to plan this when you're only looking at one element in Illustrator.}
However, two participants wanted to keep planning colors early on and saw \sysname{} more as a supplementary tool to enhance the current workflow,
\pqt{I think planning colors early is still important. I want to keep doing this, but I can see how \sysname{} could help make changes later on.}{P6}.
Thus, we believe that \sysname{} provided a promising way to make working with colors in videos more flexible and easier. 
While it is not conclusive that it will change how most people work with colors at first, it offers new and exciting possibilities for working with colors later in the process. 

\subsubsection{Suggestions and Limitations of \sysname{}}
During our study, participants pointed out a few areas where \sysname{} could potentially improve to make the color adjustment process more adaptable and user-friendly. 
Here are some of the feedback and suggestions.

\textbf{Adjusting the Scene Palette's sorting mechanism.}
At present, the sorting in the Scene Palette is based on color distances in LAB space. 
However, participants naturally felt that colors should be sorted according to their spatial relationship on each frame.
Many participants (5/12) mentioned they were expecting that colors from neighboring elements would be placed close to each other in the palette. 
One of the users noted, \pqt{I kept thinking that colors from elements close to each other would also be near each other in the palette. It feels more intuitive that way.}{P12}.

\textbf{Enhancing flexibility in color filtering.}
The current system highlights neighboring colors in the Scene Palette based on their LAB color space distance when hovering over the theme color on the Video Theme panel. 
However, participants suggested having more control over this feature could be beneficial. 
They mentioned the desire to group and select colors based on different criteria sometimes, depending on the specific adjustment task at hand. For instance, at times, they might want to group colors with the same hue but different tones, and on other occasions, they may prefer grouping colors with similar tones but different hues. 
As one participant suggested, \pqt{It would be so much more helpful if we could define the rules for grouping colors, sometimes I want to focus on hue, and other times on tone. It would allow for more precise adjustments.}{P7}.

\section{Discussion}
We reflect on the design of \sysname{} and discuss the lessons we learned and the implications for future research.

\subsection{Learned Lessons for Design}
\subsubsection{Supporting the Comprehensive Understanding of Color Distribution Across the Timeline.}
Our findings revealed that designers appreciate the multi-view (\ie, Video Theme, Scene Palette, Element List) design of \sysname{}.
These views offer varied levels of abstraction, allowing for a detailed depiction of color information at different granularities. Such visualization aids designers in forging a hierarchical comprehension of color distribution, guiding them from a general overview of dominant colors to in-depth insights into intricate color patterns, thereby facilitating a progressive color modification process.
The Scene Palette, in particular, stands as a middle-level view to display the colors across the video timeline. 
Designers noted its efficacy in offering a richer understanding of color evolution throughout the video, a distinct advantage over traditional methods sticking on static frames or individual elements.

We believe that similar techniques could be employed to edit other properties in MG videos, such as animations. 
The present authoring tools for MG videos tend to focus on individual frames, lacking a broader perspective that encompasses the entire video narrative. 
It is worth investigating exploiting such a holistic visualization strategy, possibly with varied abstraction levels to highlight different facets, to enable designers to intuitively understand how an edit (for color, motion, or other aspects of MG videos) impacts the whole video. 
This innovation promises a more contextual and comprehensive approach to MG video editing, expanding beyond the constraints of single-frame editing.

\subsubsection{Avoiding Ambiguous Color Assignment During Manipulation Through Direct Color-Element Linkage.}
Our findings highlighted a crucial, yet often overlooked, aspect in the design of interactive palettes: the necessity to establish a clear, explicit color-element association between the interactive palette and the video to ensure precise color assignment during the color manipulation phase.
Designers particularly appreciate the feature of synchronized element highlighting when adjusting color blocks within the Scene Palette. 
This preference stems from a deeper need for precise control during the color manipulation phase. 
Initially, during the color understanding phase, designers can comfortably work with abstractions, \ie, Scene Palette in \sysname{}, to intuitively perceive harmony and interplay without diving into detailed color-element connections. 
This aligns with previous studies \cite{jalal2015color}, highlighting the benefits of color abstraction for a better understanding of color interdependence. 
However, when transitioning to the color manipulation phase, designers need to know which element is being altered with a chosen color. 
As pointed out by designers, without the direct correlation, there will be a lot of ``guesswork'' that comes with guessing which colors correspond to which elements, a challenge exacerbated in complex videos hosting a lot of elements with similar colors. 
This will result in the reluctance to use the interactive palette, leading designers to resort to the more precise but tedious object-driven way, tweaking colors for each object individually.

While existing research, such as Color Triads \cite{shugrina2020nonlinear}, validated the effectiveness of interactive palettes to facilitate rapid color manipulations, they overlook such demands of precise control, thus restricting their use to the exploratory stage and failing to aid designers in producing a polished design.
We believe that incorporating the direct linkage mechanism could largely enhance the efficacy of interactive palettes, making it more effective in supporting color manipulations for both MG video and static designs by eliminating the need for guesswork and mitigating ambiguous color assignments.

\subsubsection{Reducing Navigation Burden for Locating Colors in the Complex Interactive Palette.}
During our study, we noticed that participants frequently used the Video Theme panel as a color filter, selecting the highlighted neighboring colors in the Scene Palette to finalize their choices before undertaking group-wise color modifications. 
This behavior emerged from a common tendency to modify a theme color alongside its varying neighboring colors, which might be dispersed unevenly across the Scene Palette, presenting a difference in visibility and size.
This pattern reveals that in the case of MG videos, characterized by a rich color palette and complex scenes, navigating the corresponding interactive palette and selecting multiple target colors as a group can become cumbersome. 
The integration between the Video Theme and the Scene Palette naturally alleviated this issue, facilitating smoother navigation to key colors.
However, there is still room for improvement to further minimize navigation efforts.
Currently, the neighboring color highlight operates based on a fixed, predetermined threshold distance, a parameter that could be developed to offer user-controlled adjustments, providing greater flexibility in selecting color ranges.
Moreover, introducing additional navigation aids, such as a color search functionality through user input queries, could potentially make the color navigation more smoother.
For future developments, we recommend an increased focus on facilitating smoother navigation and exploring search mechanisms to make the interactive palettes more efficient. 

% \subsubsection{Encouraging creative theme exploration by allowing flexible group-wise color manipulation.}
% The group-wise color manipulation is the special benefit brought by the interactive scene palette in \sysname{}. 
% During the study, we observed participants heavily rely on the Scene Palette to modify colors and generally start from group-wise color manipulation.
% The grouping can be both grouping similar colors for collective hue adjustments or distinct colors for saturation or lightness adjustment. 
% This indicates that allowing flexible selecting colors to form a group is necessary.
% Forming a group 
% And such grouping usually involves colors in different elements spanning across different frames, without such a color distribution visualization across the timeline, such grouping will not possible. 
\rev{
\subsubsection{Enabling Group-wise Manipulation for Design Elements Beyond Colors.}
% The feature of group-wise color manipulation in \sysname{} has proven to be particularly beneficial, highlighting the interdependence of colors throughout a video. 
% \jian{what it highlight the independence throughout the video?} 
The group-wise color manipulation feature in \sysname{} has proven to be particularly beneficial for video color design where color usage is usually considered in groups to maintain visual continuity and harmony. 
This feature makes the editing more efficient when designers need to adjust a group of relevant colors together.
This concept of grouping is not exclusive to colors, which can extend to typography, animations, stroke styles, and other design elements in MG videos.
For instance, videos usually employ consistent fonts across various scenes. 
When changes are needed, designers usually do not just modify one specific instance of text. % but an entire set. 
Similarly, to maintain visual continuity, designers might reuse certain animation types or adopt similar motion designs across different scenes. 
Grouping these animations and enabling group-wise modification across scenes can efficiently simplify the authoring process. 
However, the challenge lies in the implementation of effective grouping mechanisms. 
In \sysname{}, we introduced both auto- and manual-group options (Sec.~\ref{sec:group-wise}) for color management; similar strategies could be adapted for other design elements. 
% In addition, it is necessary to provide users with clear, immediate feedback on grouped elements. 
% This could be achieved through well-designed visual representations, allowing for quick and easy identification and assessment of grouped elements. 
% \jian{a bit vague on what visual representations}
Moreover, given that target elements can span across multiple scenes, it is necessary to offer a clear and comprehensive overview of all selectable elements for grouping, along with instant feedback on the grouped ones for easy identification and evaluation. 
A feasible solution is to design a proper visual abstraction for these elements and display them on a timeline, for example, the Scene Palette in \sysname{}.
Such an approach would allow users to view all elements together, enabling efficient selection without navigating through different frames. 
The immediate visual feedback (\eg, using the black outline highlighting selected colors in \sysname{}) could ensure the clear tracking of which elements have been grouped. 
Future work could explore the tailored design principles for applying the group-wise editing to other design aspects such as animation authoring. 
}

\subsection{Future Work}
% \jian{think how you can combine limitation and future work}

\subsubsection{Exploring the Design Space of the Color Visualization for MG Videos.}
% spatial information in the scene palette;
% frame rate, time step;
% transition;
% color gradient;
In this project, our primary aim was to develop a holistic system for context-aware, group-wise, and multi-level MG video color authoring, with the interactive palette as a crucial component. We did not extensively explore the design space for the Scene Palette nor did we evaluate alternative possible designs.
What we learned at this stage suggests empirical evidence for the utility of using such interactive palettes in exploratory video color authoring.
The existing design decisions around the Scene Palette were influenced both by prior studies emphasizing the significance of color proportions \cite{wyszecki2000color, whitfield1990color}, and expert designers' input.
Nonetheless, while current encoding mechanisms are reasonable, there may yet be room to expand upon the complexity of color distribution visualization within MG videos. For example, future work could explore the role of spatial color information within each frame, \rev{investigate advanced color sorting algorithms to ensure consistent color positioning across frames,} and examine how different design alternatives of Scene Palette impact designers' perception towards color usage and usage patterns.

\subsubsection{Providing Flexible Parametric Configuration of Palette Visualization.}
In recognizing that each designer and project exhibits unique needs, we envision the potential for future adjustments to the palette designs (both the Video Theme and Scene Palette views). 
We understand that there isn't a ``one-size-fits-all'' palette design and it would be helpful to align the palette design with specific usage scenarios and video characteristics. 
\rev{
The Video Theme view currently displays five colors, but making this number flexible would more effectively accommodate the diverse complexities of motion graphic videos and users' desired levels of control.
}
In addition, it would also be beneficial to enable users to adjust the Scene Palette's parametric configuration, such as the time step and color sorting. 
We believe that by tailoring palette visualization to individual requirements, we can better serve the diverse needs and intentions of video creators, driving the progression of color authoring for MG videos.

\subsubsection{Supporting Diverse Asset Types.}
Our current solution depends on obtaining color information directly from the structural data of SVG elements. 
However, not all MG videos solely utilize SVG format elements; some may also include raster images or GIF videos to augment expressiveness. 
Extracting color data and corresponding properties, such as proportion from these rasterized elements, is not straightforward and generally requires specific algorithms. 
Therefore, dealing with mixed element formats and subsequently integrating the color data into a palette for comprehensive video color distribution presents a challenge. 
Addressing this issue and finding ways to handle all element formats uniformly is crucial, and is an open question we aim to explore in our future work.

\rev{
\subsubsection{Rethinking the Workflow of Motion Graphics Creation with \sysname{} Integrating into After Effects.}
\sysname{} is currently developed as a standalone application, and our study design focuses on the post-animation stage color modifications.
The in-lab user study validates the utility of \sysname{} in late-stage color authoring and reveals the advantages of contextualizing color design within an established narrative flow.
However, we are also aware that, in the real world, design is an iterative rather than a linear process. 
Color authoring is more likely to span the entire video creation process and be intermingled with other modifications made to the graphical elements or to their animations.
We envision that integrating \sysname{} into existing animation tools (\eg, Adobe After Effects) could offer the flexibility to enhance such an iterative process, allowing designers to effortlessly switch between color modifications and animation designs.
Furthermore, such integration would likely encourage a deeper exploration of the interplay between colors and animations.
For instance, if a designer decides to brighten a group of colors for a more vibrant atmosphere, they might also instinctively adjust the corresponding motions to amplify this sense of excitement, and similarly, changes in motion could prompt corresponding adjustments in color.  
Our future work aims to study how to better integrate \sysname{} into existing animation tools, and investigate how this integration might influence user interaction patterns, workflows, and mental models in the MG video design process.

}

\subsection{Limitations}
Our study comes with certain limitations that warrant mentioning. Firstly, we only involved expert users in the formative study and evaluation study, which does not reflect the experiences of novices. 
Therefore, our results may not reveal the challenges that beginners might face when using the color authoring tool for MG videos and the experiences of novices when using our system.
Secondly, the MG video used in our study was relatively simple, featuring just four central scenes. 
Actual MG videos can be far more complex than this. 
Hence, our findings and implications may not represent all potential scenarios and needs for a more diverse and complex set of MG videos.
Various aspects, such as video genre, scene complexity, color dynamics, or the target audience, can all influence user perceptions and requirements for color authoring tools. 
Therefore, future work should expand to include a broader range of video types, complexities, and diverse demographics to ensure the effectiveness and applicability of our tool to a wider pool of users and needs.

\section{Conclusion}
In conclusion, this research systematically explored the current process of color design for MG videos, identified the challenges in MG video color authoring, and pointed out the design opportunities to improve the authoring experience. 
Addressing this, we introduced \sysname{}, an interactive color authoring tool that simplifies color changes in MG videos, helping designers make easy and quick adjustments without losing sight of the bigger picture, and motivating the creative color theme exploration.
The expert evaluation highlights its effectiveness in supporting intuitive, progressive, and \rev{group-wise} color adjustments.
Participants also suggest the possibility of \sysname{} to work well with existing tools like Adobe After Effects, offering a smoother and more integrated workflow for designers. 
Moving forward, we believe that \sysname{} can steer further research and developments in this field, making the process of creating MG videos more streamlined and effective.

\begin{acks}
This work was completed during the internships of Xinyu Shi and Yinghou Wang at Microsoft Research Asia (MSRA). We thank all our participants for their time and valuable input. We would also like to thank our reviewers whose insightful comments have led to a great improvement of this paper. This work is supported in part by the NSERC Discovery Grant.
\end{acks}

\bibliographystyle{ACM-Reference-Format}
\bibliography{references.bib}

\end{document}